\documentclass{aa}

\usepackage{graphicx}
\usepackage{caption}
\usepackage{subcaption}
\usepackage{booktabs}
\usepackage{pgfplotstable}
\usepackage{colortbl}
\usepackage{xcolor}
\usepackage{booktabs} 
\pgfplotstableset{
    every head row/.style={
    before row=\toprule,after row=\midrule},
    every last row/.style={
    after row=\bottomrule},
    col sep = &,
    row sep=\\,
    string type,
}
\usepackage{txfonts}
\usepackage{float}
\usepackage{natbib,twoopt}
\usepackage{soul} 
\usepackage[breaklinks=true]{hyperref} 
\bibpunct{(}{)}{;}{a}{}{,}             
\makeatletter
  \newcommandtwoopt{\citeads}[3][][]{\href{http://adsabs.harvard.edu/abs/#3}%
    {\def\hyper@linkstart##1##2{}%
     \let\hyper@linkend\@empty\citealp[#1][#2]{#3}}}
  \newcommandtwoopt{\citepads}[3][][]{\href{http://adsabs.harvard.edu/abs/#3}%
    {\def\hyper@linkstart##1##2{}%
     \let\hyper@linkend\@empty\citep[#1][#2]{#3}}}
  \newcommandtwoopt{\citetads}[3][][]{\href{http://adsabs.harvard.edu/abs/#3}%
    {\def\hyper@linkstart##1##2{}%
     \let\hyper@linkend\@empty\citet[#1][#2]{#3}}}
  \newcommandtwoopt{\citeyearads}[3][][]%
    {\href{http://adsabs.harvard.edu/abs/#3}
    {\def\hyper@linkstart##1##2{}%
     \let\hyper@linkend\@empty\citeyear[#1][#2]{#3}}}
\makeatother
\begin{document}

   \title{Kinematic evidence of magnetospheric accretion for Herbig Ae stars with JWST NIRSpec}


   \author{Ciarán Rogers
          \inst{1},
          Bernhard Brandl,
          \inst{1, 3}
          Guido de Marchi
          \inst{2}}

   \institute{Leiden Observatory, Leiden University,
              PO Box 9513, 2300 RA Leiden, The Netherlands\\
              \email{rogers@strw.leidenuniv.nl}
         \and
             European Space Research and Technology Centre, Keplerlaan 1, 2200 AG Noordwijk, The Netherlands\\
             \email{gdemarchi@rssd.esa.int}
         \and
            Faculty of Aerospace Engineering, Delft University of Technology, Kluyverweg 1, 2629 HS Delft, The Netherlands\\
            \email{brandl@strw.leidenuniv.nl}}

  \date{Received December 9, 2024; accepted April 18, 2025}
  \titlerunning{Line kinematics for Herbig AeBe stars}
  \authorrunning{Ciarán Rogers, Bernhard Brandl, Guido de Marchi}

 
  \abstract
   {Hydrogen emission lines have been used to estimate the mass accretion rate of pre-main-sequence stars for over $25$ years. Despite the clear correlation between the accretion luminosity of a star and hydrogen line luminosities, the physical origin of these lines is still unclear. Magnetospheric accretion (MA) and magneto-centrifugal winds are the two most often invoked mechanisms.}
   {Using a combination of HST photometry and new JWST NIRSpec spectra in the range $1.66 - 3.2 \; \mu m$, we analysed the spectral energy distributions (SEDs) and emission line spectra of five sources in order to determine their underlying photospheric properties and attempt to reveal the physical origin of their hydrogen emission lines. These sources reside in NGC 3603, a Galactic massive star forming region.}
   {We performed fits of the SEDs of the five sources employing a Markov chain Monte Carlo exploration to estimate $T_{eff}$, $R_{*}$, $M_{*}$, and $A(V)$ for each source. We performed a kinematic analysis across three spectral series of hydrogen lines (Paschen, Brackett, and Pfund), totalling $\ge 15$ lines per source. We studied the full width at half maximum (FWHM) and optical depth of the spectrally resolved lines in order to constrain the emission origin. We calculated the expected velocities from MA as well as gas in Keplerian orbit for our sources.}
   {All five sources have SEDs consistent with young intermediate-mass stars. We classified three of these sources as Herbig Ae type stars based on their $T_{eff}$. Their hydrogen lines show broad profiles with FWHMs $\ge 200$ km s$^{-1}$. Hydrogen lines with high upper energy levels $n_{up}$ tend to be significantly broader than lines with a lower $n_{up}$. The optical depth of the emission lines is also highest for the high-velocity component of each line, and it becomes optically thin in the low-velocity component. Three sources show FWHMs that are too broad to originate from Keplerian rotation, but they are consistent with MA. The remaining two sources have FWHMs that are consistent with both MA and Keplerian rotation.}
   {The highest excitation lines have the largest FWHM for a given series. The highest-velocity component of the lines is also the most optically thick. This is consistent with emission from MA or a Keplerian disc, but it cannot be explained as originating in a magneto-centrifugal wind. Based on the expected velocities from MA and a Keplerian disc, we favour MA for three of the five sources. We cannot rule out Keplerian disc emission for the remaining two sources. In the future, this approach can be applied to more statistically significant samples of Herbig AeBe spectra, including existing archival observations.}


   \maketitle
%

\section{Introduction}
\label{sec:intro}
Protoplanetary discs around young stars are the site of numerous highly studied astronomical phenomena, including magnetospheric accretion and accelerating disc winds powered by magneto-centrifugal forces, and as the name suggests, they are the formation sites of planets. Understanding the evolution of protoplanetary discs is crucial in both the context of star formation and planet formation. A fundamental push and pull that regulates the lifetime of protoplanetary discs is the balance between mass-loss through winds and mass accretion onto the central star.\\
Accretion rates of stars were originally measured by fitting shock models to the near-ultraviolet (NUV) spectra of young low-mass stars, known as Classical T Tauri Stars (CTTSs) \citep{gullbring1998disk}. The physical picture assumed here is centred around a magnetically driven accretion paradigm in which the star's strong magnetic field pushes against and truncates the protoplanetary disc out to a few stellar radii \citep[e.g.][]{hartmann2016accretion}. Material at the disc's surface is channelled along magnetic field lines from the disc to the central star. During its near free-fall towards the star, the supersonic material shocks the stellar surface and releases its gravitational potential energy as radiation. Initially this is primarily in the form of X-rays, which are quickly absorbed by the stellar surface and re-radiated at longer wavelengths in the NUV and blue-optical. This additional source of NUV light on top of the intrinsic photospheric contribution can be observed and measured, and from this so-called accretion luminosity, the mass accretion rate can be determined. This is the only direct method of measuring the accretion luminosity, and while it unambiguously probes the accretion shock, the reliance on NUV wavelengths makes this approach impractical for sources that are deeply embedded or located at great distances due to high levels of extinction. \\
An invaluable calibration was discovered by \cite{muzerolle1998brgamma}, who showed that the line luminosity of the relatively bright near-infrared (NIR) hydrogen line $Br_{7}$ scales tightly with the accretion luminosity for a sample of $19$ CTTSs. With extinction more than ten times lower in the NIR compared to the NUV and without the need for shock modelling, this line luminosity calibration opened the floodgates for the efficient and straightforward estimate of accretion rates for large samples of sources. These calibrations have since been updated and expanded, most notably by \cite{herczeg2008uv, alcala2014x, alcala2017x}. We also showed in an earlier work that these calibrations could be extended towards the brightest NIR line - $Pa_{\alpha}$ \citep{rogers2024determining}, thus making estimating the accretion rate highly accessible in the era of JWST for extremely distant and embedded sources.\\
All of this work relies on the assumption that accretion is driven magnetically. First theorised by \cite{koenigl1991disk} and unambiguously confirmed observationally by \cite{bouvier2007magnetospheric} for CTTSs, the MA paradigm for Herbig AeBe stars has not reached the same consensus. Unlike CTTSs (spectral types M through G), stars of spectral type A and earlier are not expected to possess a convective envelope, which is thought to be the source of the strong and ordered magnetic fields present around a CTTS \citep[e.g.][]{kageyama1997generation, johns2007magnetic}. Despite lingering questions over their origin, magnetic fields have been measured towards dozens of Herbig AeBe stars, albeit with significantly lower field strengths compared to CTTSs \citep{mendigutia2020mass}. Other spectroscopic signatures of infalling matter have also been seen towards Herbig AeBe stars, such as red-shifted absorption profiles in He I and H I lines \citep{cauley2015optical}, though they are less prevalent compared to the occurrence rate in CTTSs. In general, magnetospheres, if present, are expected to be smaller in Herbig Ae stars compared to CTTSs and may be entirely absent for early Be stars \citep{wichittanakom2020accretion, vioque2022identification}.\\
Disc winds are ubiquitous around young stars. However, as with accretion, less is known about winds from Herbig AeBe stars compared to CTTSs. Powerful collimated jets and outflows have been spatially resolved for many nearby sources, with large-scale jets being the defining physical feature of the class of objects known as Herbig-Haro objects \citep{reipurth2001herbig}. The exact launching mechanism of these winds is still under debate, although there is a growing consensus that these outflows are the result of a magneto-centrifugal wind -- also referred to as a magnetohydrodynamical (MHD) wind \citep[e.g.][]{tabone2022mhd} -- first theorised by \cite{blandford1982hydromagnetic}. These winds are expected to be crucial in the removal of angular momentum from the disc and are theorised to ultimately facilitate accretion by ejecting material, causing outer disc material to migrate inwards and replenish the inner gas disc where accretion takes place \citep{frank2014jets}. Studies focusing on disc winds from Herbig AeBe stars, however, have presented contradictory results. In \cite{kraus2008origin}, the authors used high spatial resolution interferometric observations of the $Br_{7}$ line to infer the emission mechanism. They found that four out of five sources were consistent with emission from a stellar or disc wind, with the final source showing more compact $Br_{7}$ emission more consistent with emission from the MA flow. In \cite{kreplin2018brgamma}, using a similar approach as \cite{kraus2008origin}, the authors also found that $Br_{7}$ emission from the Herbig AeBe star MWC 120 was consistent with a disc wind. In \cite{tambovtseva2016brackett}, the authors used non-local thermodynamic equilibrium modelling of $Br_{7}$ emission from a Herbig Ae star and also found that the $Br_{7}$ emission likely originates in a disc wind, not an MA flow. In contrast to the above studies, \cite{cauley2014diagnosing} used the metastable transition of HeI $\lambda \; 10830 \; \AA$ as a tracer of mass infall and outflow, which is commonly used for similar purposes with CTTSs \citep[e.g.][]{2005ApJ...625L.131D, edwards2006probing}. In their sample of $56$ Herbig AeBe stars, the authors found no evidence of narrow blue-shifted absorption, which traces outflowing material due to disc winds. They did however find red-shifted absorption tracing infalling material for the Herbig Ae stars in their sample, which is indicative of MA. The same red-shifted absorption was not seen towards the Herbig Be stars in their sample. These results do not paint a clear or consistent picture, and they highlight the need for further studies to investigate the accretion and outflow properties of Herbig AeBe stars.\\ 
There is no doubt that both winds and accretion give rise to hydrogen emission lines, but there is not yet a consensus on which of the two processes dominates the hydrogen line emission for a given stellar mass and stage of evolution. Using the JWST NIRSpec Micro-Shutter Assembly (MSA) \citep{ferruit2022near}, we obtained 100 stellar spectra, including a number of young intermediate mass stars and Herbig Ae stars. These stars all reside in the giant Galactic star forming region NGC 3603 located $7.2 \pm 0.1$ kpc away \citep{drew2019star}. We present five of these sources here, which all show rich emission line spectra featuring large unbroken sections of the Brackett and Pfund hydrogen series as well as the strong $Pa_{\alpha}$ line. We performed a kinematic analysis on the hydrogen lines for each source in order to constrain the physical origin of these lines. \\
In Section ~\ref{sec:sel_of_tar}, we discuss the target selection. In Section ~\ref{sec:data_reduction}, the data reduction and post-processing steps are explained and the NIRSpec spectra are shown. In Section ~\ref{sec:sed_fitting}, we present the spectral energy distribution (SED) fitting of our sources. In Section \ref{sec:measure_lines}, we describe our methods for fitting and measuring the emission lines. In Section ~\ref{sec:results}, the kinematic analysis is presented. In Section ~\ref{sec:discussion}, we interpret our analysis and suggest a physical origin for the emission lines. We also address alternative emission line mechanisms outside magnetic accretion and winds. In Section ~\ref{sec:conclusions}, we summarise our findings.

\section{Observations and target selection}
\label{sec:sel_of_tar}
\subsection{NIRSpec observations} \label{subsec:nirspec_obs}
The target selection procedure has already been described in \cite{rogers2024determining} and \cite{RogersCTTS}. Here we summarise it briefly with a specific focus on the five sources presented in this study. 100 stellar spectra were obtained as part of a NIRSpec Guaranteed Time Observations (GTO) programme (PID=1225, PI: De Marchi, 2022). The aim of this program was to obtain NIRSpec spectra of stars that showed strong line emission due to accretion related activity, as inferred from photometric $H_{\alpha}$ measurements (see \cite{beccari2010progressive}). The five targets presented here represent the brightest, most massive and strongest accreting of the sample. These sources had not been studied in detail before, and we did not know prior to these new NIRSpec observations that they constituted a sample of intermediate mass T Tauri stars and Herbig Ae stars (see Section \ref{sec:sed_fitting}). As such, we did not set out to obtain spectra of these sources in order to perform a kinematic analysis of their hydrogen lines. Rather, upon inspection of their rich emission line spectra, we were naturally motivated to analyse them further. The SIMBAD names and coordinates of these sources are shown in table \ref{herbigs_coords}.\\
Our observing strategy consisted of opening three neighbouring micro-shutters for each source in a column of the MSA, forming a `mini-slit'. Each source was placed in the central shutter, with the upper and lower micro-shutters observing the surrounding nebula. We employed a nodding pattern, nodding the telescope three times such that the source moved from the central shutter, to the upper shutter and finally the lower shutter. This pattern provided us with three exposures for each source, which could then be averaged together. We found in our analysis that combining exposures to create an average spectrum could increase the measured full width at half maximum (FWHM) of the emission lines by up to $50 km \: s^{-1}$. This may be due to slight differences in the wavelength solution between the three exposures, leading to smearing of the emission lines after averaging. Because of this, we opted to instead work on single exposures. For these five bright sources, this did not present any problems as the marginal boost in S/N due to averaging was not needed. The sources were observed with the grating/filter combination F170LP/G235H, providing us with wavelength coverage from $1.66 \: - \: 3.2 \: \mu m$, with a typical resolving power of $\sim4000$ (the subject of NIRSpec's spectral resolution is discussed in \ref{subsec:spec_res}.
\subsection{HST observations} \label{subsec:HST_obs}
As part of our analysis we also included optical and NIR HST/WFC3 photometry (PID: 11360, PI: O'Connell, 2009). The filters employed from this program were F555W, F625W, F814W, F110W and F160W. A photometric study of the pre-main-sequence population of NGC 3603 was presented in \cite{beccari2010progressive} using the F555W and F814W filters. Our NIRSpec observations serve as a spectroscopic follow up to those observations. We also used the F435W filter from HST/ACS observations (PID: 10602, PI: Maiz-Apellaniz, 2005). Both programs observed the centre of NGC 3603 with fields of view ranging from $123 \arcsec\times139 \: \arcsec$ for the WFC3 IR channel, $160 \arcsec\times160 \: \arcsec$ for WFC3 UVIS channel, and $202 \arcsec\times202 \: \arcsec$ for the ACS. These two programs were not contemporaneous, taking place almost five years apart. This introduces the potential for photometric variability. Likewise, our NIRSpec spectra were not observed contemporaneously with either of the HST programs. In \cite{RogersCTTS}, we flux calibrated our NIRSpec spectra using the F160W photometry of each source, after performing a small extrapolation to cover the wavelength gap. The NIRSpec spectra were in excellent agreement with F160W photometry, with typical flux correction factors of $1.02 \: \pm \: 0.11$. This is in agreement with the expected flux accuracy of NIRSpec \cite{gordon2022james}, strongly implying that variability does not significantly affect the continuum/broadband photometry for these non-contemporaneous observations. We proceeded under this assumption for the remainder of the analysis.

\begin{table}
\caption{SIMBAD names and coordinates of each source.}
\begin{tabular}{@{} l *3c @{}}
\toprule
 \multicolumn{1}{c}{ID} & SIMBAD ID & RA & DEC \\ 
\midrule
 185 &  [SB2004] 56227 & 168.762769 & -61.262512\\
 238 &  [SB2004] 57135 &  168.815242 & -61.255092\\
 251 &  2MASS J11150443-6115112 & 168.768481 & -61.253132\\
 469 &  [FPA77] NGC 3603 IRS 9B & 168.794205 & -61.277950 \\
 823 &  [SB2004] 56216 & 168.762378 & -61.271019 \\ \bottomrule
 \end{tabular}
 \label{herbigs_coords}
\end{table}

\section{Data reduction}
\label{sec:data_reduction}
\subsection{NIPS}
The observations were largely reduced with the ESA Instrument Team’s pipeline known as the NIRSpec Instrument Pipeline Software (NIPS) \citep{NIPS}. Additional reduction steps were also written specifically for these observations which are briefly discussed. NIPS is a framework for spectral extraction of NIRSpec data from the count-rate maps, performing all major reduction steps from dark current and bias subtraction to flat fielding, wavelength and flux calibration, background subtraction and extraction, with the final product being the 1D extracted spectrum.\\
One of the final steps before extraction is the rectification of the spectrum. The dispersed NIRSpec spectra are curved along the detector. Rectification is performed in order to `straighten' the spectrum. By doing this, the spectrum is resampled onto a uniform wavelength grid (each wavelength bin $\Delta \lambda$ is the same for every pixel). 
The rectified spectrum is a count-rate map consisting of 3817 pixels in the dispersion direction and 7 pixels in the spatial direction. The final data reduction step - extraction, simply collapses the 2D spectrum by summing the 7 pixels along each column.
\subsection{Nebular background subtraction}
\label{subsec:neb_sub}
\begin{figure*}[h]
    \centering
    \includegraphics[width=0.75\linewidth]{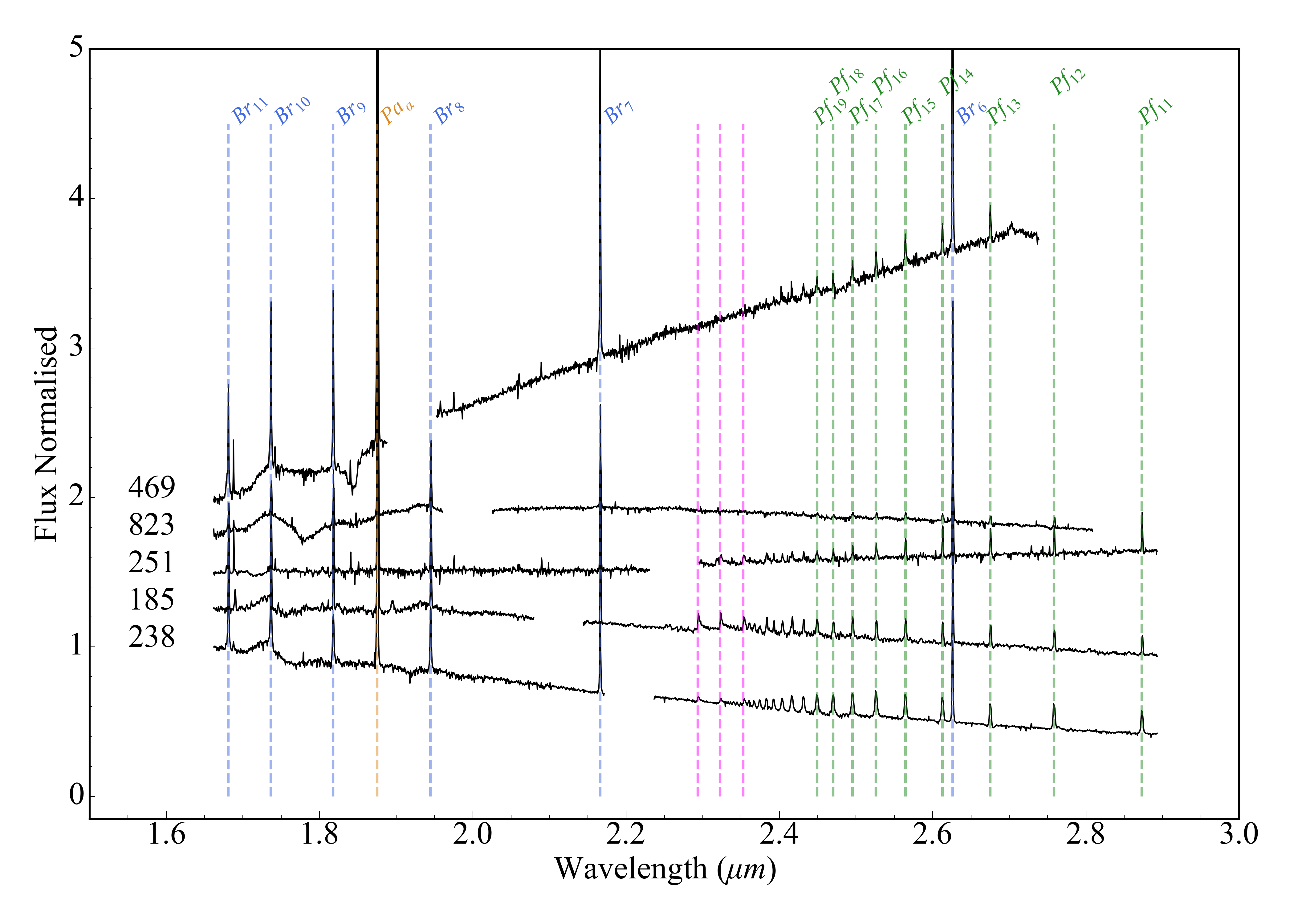}
    \caption{NIRSpec spectra of our five PMS sources. The spectra have been normalised at $1.66 \mu m$. For better visualisation, they have been offset from each other vertically and cut off after the final Pfund line at $\sim 2.9$ $\mu m$. Source names are indicated on the left of the spectra. Hydrogen emission lines are marked with vertical dashed lines; the Paschen series is in orange, the Brackett series is in blue, and the Pfund series is in green. The CO bandheads are also indicated with pink dashed lines. A systematic calibration uncertainty impacts the continuum at $\sim 1.7\mu m$}
    \label{fig:Herbig_spectra}
\end{figure*}

The bright hydrogen emission lines originating from the H II region were subtracted from each of the stellar spectra following the approach outlined in detail in \cite{RogersCTTS}. This approach makes use of the nebular He I line at ($\lambda$ = $1.869 \mu m$). The nebular spectrum of each source was scaled and then subtracted, such that the He I line was completely removed from the stellar spectrum. We have estimated that the uncertainty this introduces to the final flux of the hydrogen lines is $\sim 10 \%$. Figure \ref{fig:Herbig_spectra} shows the subtracted NIRSpec spectra of the five sources.
\subsection{Spectral resolution of NIRSpec MSA}
\label{subsec:spec_res}
The often quoted resolving power for NIRSpec in high resolution mode is $\frac{\lambda}{\Delta\lambda} \sim 2700$. This is based on pre-flight calculations, assuming a fully illuminated micro-shutter, with each resolution element being sampled by $2.2$ pixels on the NIRSpec detectors. For point sources however, the resolving power increases by a factor of roughly $\sim 1.5$. To explain this effect, the connection between a micro-shutter and the pixels onto which it projects needs to be understood. The micro-shutters have been designed to project geometrically over two detector pixels, ensuring that for a fully illuminated micro-shutter the line spread function (LSF) is Nyquist sampled \citep{ferruit2022near}. A uniformly illuminated micro-shutter can be imagined as containing an infinite series of point sources. Each of these point sources projects to a slightly different region of the detector. When light is dispersed from each of these point sources, the LSFs of neighbouring points blend together, forming a broader LSF that is composed of many intrinsically narrower LSFs. This blending reduces the fundamental spectral resolution that is achievable with NIRSpec. In the case of a point source, this blending does not occur, as there is only one source of light projecting onto the detector. This means that a higher spectral resolution is achievable when observing point sources compared to extended sources. This has the side effect that, for point sources, the NIRSpec LSF is not Nyquist sampled at any wavelength.\\
Given that all of our sources are point sources, the actual spectral resolution achieved for our observations needed to be determined if we wished to analyse the kinematics of the emission lines. We have used the JWST-MSAfit software from \cite{de2024ionised} to do this. This software provides forward modelling and fitting of simulated MSA data. It allows the user to first select a grating and filter combination, and then define a source (extended or point-like) and place it in a specified micro-shutter at a specified position within that micro-shutter. A spectrum with $17$ uniformly spaced emission lines is generated from this. The $FWHM$ of the emission lines are measured in order to determine what the resulting resolving power is at each wavelength. As already mentioned, the LSF is not Nyquist sampled by the NIRSpec detector in high resolution mode for point sources. This results in a somewhat jagged FWHM curve (see black line in figure \ref{fig:238_FWHM}), as some line peaks are centred on one pixel, while others are spread over two. The estimated uncertainty of this approach is expected to be between $10-20\%$ \citep{de2024ionised}. Using this tool we have determined that the actual resolving power for our sources is $\frac{\lambda}{\Delta\lambda} \sim 4300$, at $\lambda = 1.875 \mu m$. This corresponds to a FWHM of $\sim 70 \pm 14 km s^{-1}$, assuming an uncertainty of 20 \%. 

\section{Determining the stellar properties of our sources}
\label{sec:sed_fitting}
In order to draw physical conclusions from the emission lines of our sources, the physical properties of the stars needed to be determined. Given that none of the five sources display photospheric absorption lines in their NIRSpec spectra, traditional spectral classification was not possible. Instead we relied upon SED fitting to determine the underlying photospheric properties and extinction of each source. To do this, we have used archival Hubble Space Telescope (HST) photometric observations in order to perform aperture photometry and measure the optical and infrared fluxes of the sources. The filters employed were F435W (ACS), F555W, F625W, F814W, F110W, and F160W (WFC3). We explored the archives for other photometric observations of these sources, but given the distance and the highly crowded field of NGC 3603, we were forced to rely solely on HST. These observations are the only available with the necessary spatial resolution and sensitivity to observe these sources with high S/N, and without confusion in the beam from other nearby stars. The stellar photosphere is well traced by optical photometry as the star's light dominates at wavelengths between $0.4$ and $0.8 \; \mu m$ \citep{bell2013pre}. This is especially for true for Herbig stars whose hotter photospheres peak at $\sim 0.5\;\mu m$. We have used a Markov Chain Monte Carlo exploration to fit the photometry of each source, determine the best-fitting stellar parameters, and estimate uncertainties on those parameter values.\\

\subsection{Indirect clues about the spectral type of the continuum sources} \label{subsec:indirect_clues}
There were some immediate clues about the underlying stellar properties of our continuum sources. They are highly luminous, representing the highest S/N sources in the sample, and show relatively flat, or rising SEDs (see Section 6.2 of \cite{RogersCTTS} for a discussion on the spectral index of the sources). The high luminosities and lack of absorption features in their spectra is naturally explained if these sources are of intermediate or high mass, with relatively hot photospheres. On the other hand, if these sources were actually low-mass CTTS, the high luminosities could possibly be explained as a burst of accretion, with the lack of absorption features in their spectra being a result of exceptionally high levels of excess veiling emission from the protoplanetary disc. Veiling emission is often modelled as a blackbody \citep[e.g.][]{muzerolle2003unveiling, cieza2005evidence, mcclure2013characterizing, antoniucci2017high, alcala2021giarps}. We have simulated how much veiling emission would be needed in order to render the strong metal absorption feature Mg I $\lambda 1.71$ $\mu m$ completely undetectable for a 4000K star with $[Fe/H]= 0.0$ and $log(g) = 4.00$, at the typical S/N of our continuum sources $S/N \sim 100$. We found that to render this line undetectable, a veiling factor $r_{\lambda}$ of $\frac{F_{disk}}{F_{star}} \ge 20$ was required. Such a high veiling factor also dramatically changes the shape of the SED in the NIR, becoming dominated by the shape of underlying blackbody spectrum. We found that at such high levels of veiling there was no combination of veiling, effective temperature ($T_{eff}$), and extinction $A(V)$ that could match our observations. From this, a simpler explanation emerges that these continuum sources simply have hot photospheres that intrinsically lack strong metal absorption lines, and are dominated by hydrogen absorption. These hydrogen absorption lines are entirely filled in by accretion related emission, leaving the spectrum with no detectable absorption lines. We tested this by fitting the SED of our sources in order to determine their $T_{eff}$.
\subsection{Fitting only optical wavelengths}
We initially attempted to fit the optical and NIR observations from HST and JWST in order to constrain the photospheric ($T_{eff}$), extinction ($A(V)$) and disc ($T_{veil}$, $r_{\lambda}$) properties simultaneously. We ultimately found that the observations available for these sources were not well suited for such a detailed analysis, and by attempting to fit both stellar and disc features, we introduced too much flexibility for our limited wavelength coverage. We lack observations longwards of $\sim 3\;\mu m$ which would constrain the shape of the disc's SED, and break the degeneracy that we found among our fits. As such, we opted to only fit the optical photometry from HST in order to constrain $A(V)$ and $T_{eff}$. By considering only optical wavelengths, we filtered out the majority of emission from the disc, greatly simplifying the analysis (while also returning less physical information). It was still possible to determine $r_{\lambda}$ once $T_{eff}$ was known by comparing the flux of the observed spectrum to the flux of the best-fitting photospheric spectrum. We placed an additional constraint during the fitting procedure that the best-fitting photospheric model must have IR fluxes equal to or below the IR HST photometry fluxes and all NIRSpec fluxes. The flat/rising NIR SEDs of these sources are consistent with moderate to strong excess veiling emission. As such, the model photospheric spectrum should be below the observed spectrum.\\
We employed a Markov chain Monte Carlo (MCMC) exploration with the python package emcee to determine the best-fitting photospheric model for each source. We have described this procedure for all sources in \cite{RogersCTTS}. We summarise it here. MCMC is an iterative sampling method to fit models to data. It works by generating a model based on the input parameters and randomly sampling values for each parameter from a pre-defined distribution (the priors). Each time a new model spectrum is generated, it is fitted to the observed spectrum, and the log-likelihood is calculated and taken as a goodness of fit metric. If the current model provides a better fit than the previous model, then a new model is generated using the current parameter values as a starting point, plus a small random input. These small random steps in parameter space are accomplished by `walkers'. If the current model is worse than before, then the next model is generated using the previous parameter values as the starting point, again, plus a small random input. This is repeated thousands of times until a reasonable range of parameter values has been determined, and no further exploration of the parameter space can improve the fit. This is known as convergence.\\
We have used the Phoenix stellar models from \cite{husser2013new} in this analysis. We degraded each model to the spectral resolution of NIRSpec, and converted the optical portion of each spectrum into photometric data points by multiplying by the HST filter throughput curves. The MCMC requires that arbitrarily small step sizes can be taken in parameter space when generating a new model to fit the observations. To enable this, we needed to interpolate the Phoenix model spectra. We used the radial basis function interpolation method from SciPy to do this.\\ We fixed the metallicity at $[Fe/H]= 0.0$ and $log(g)=4.00$ in order to reduce the number of free parameters and hence the flexibility of the fit. We extinguished our sources using the extinction curve from \cite{gordon2023one} with a total-to-selective extinction ratio of $R(V) \;=\; 4.8$ (see \cite{rogers2024spectral} for a full discussion of the extinction properties of NGC 3603). The parameters that we allowed to vary are $T_{eff}$, $A(V)$ and a luminosity scaling term. Given that these sources have not been studied in detail before, we opted for flat (uninformative) priors. We allowed $T_{eff}$ to vary between $3000 - 12000\;K$, $A(V)$ to vary between $0-10\;mag$ and the luminosity scaling parameter to vary between $10^{-23} - 10^{-22}$.\\
We achieved good fits for our continuum sources, and as we expected from the indirect clues given above, the best-fitting $T_{eff}$ values tend to be relatively hot, with a median temperature of 7553\,K. The best-fitting parameters for our sources are shown in table \ref{MCMC_herbigs}. The best-fitting model spectrum is shown for source 238 in figure \ref{fig:best_fitting_model_238}. The best fits for the remaining sources are shown in appendix \ref{best_fitting_plots}. 

\setlength{\tabcolsep}{2pt} 
\begin{table}
\caption{Stellar parameters for each source determined through SED fitting.}
\begin{tabular}{@{} l *5c @{}}
\toprule
 \multicolumn{1}{c}{ID}    &$T_{eff}$ (K)& $M_{*}$ ($M_{\odot}$) & $R_{*}$ ($R_{\odot}$)& $A(V)$ (Mag) & $r_{\lambda}$  \\ 
\midrule
 185&  $7553^{+603}_{-607}$ &  6.78 $\pm$ 0.8 & $15.35 \pm 3.34$& $5.821^{+0.28}_{-0.334}$ & $3.6\pm0.48$\\
 238&  $9893^{+656}_{-477}$ &  $7.06 \pm 0.61$ & $12.27 \pm 1.94$ & $6.883^{+0.109}_{-0.114}$ & $0.73\pm0.13$ \\
 251&  $6282^{+792}_{-441}$ &  6.12 $\pm$ 1.0 & $17.18 \pm 6.83$& $5.668^{+0.579}_{-0.374}$ & $5.91\pm0.31$\\
 469&  $5276^{+218}_{-345}$ &  4.24 $\pm$ 0.53 & $9.96 \pm 1.89$& $4.831^{+0.18}_{-0.338}$ & $12.9\pm1.06$\\
 823&  $8967^{+352}_{-379}$ &  5.54 $\pm$ 0.31 & $9.46 \pm 1.0$ & $8.087^{+0.098}_{-0.138}$ & $5.65\pm0.29$\\ \bottomrule
 \end{tabular}
 \label{MCMC_herbigs}
\end{table}
\setlength{\tabcolsep}{4pt} 

\begin{figure*}[h!]
    \centering
    \includegraphics[width=1\linewidth]{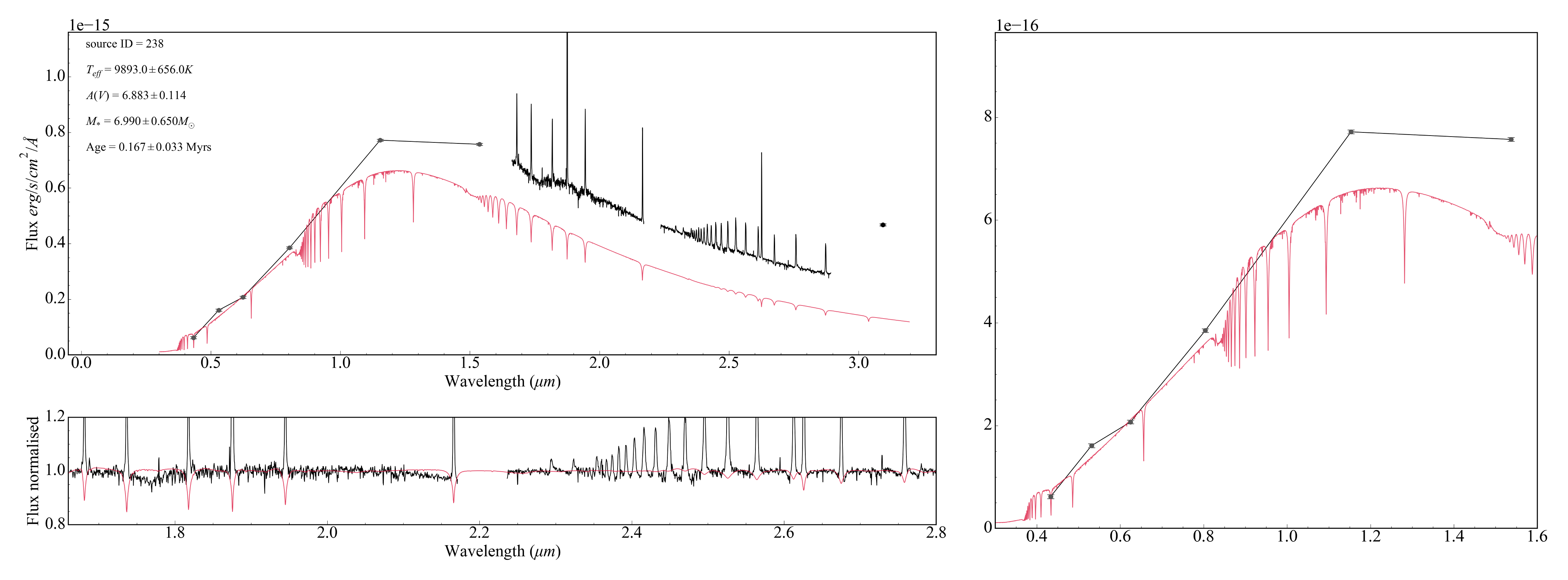}
    \caption{Top-left panel - SED of source 238 from 0.4-3 $\mu m$. NIRSpec spectrum, with uncertainty represented by error bar (black). HST photometry (grey). Phoenix stellar model (red). Bottom-left panel - Normalised NIRSpec spectrum (black). Phoenix stellar model (red). Best-fitting model shows no metal absorption lines, consistent with the NIRSpec spectrum. All hydrogen absorption lines are filled in by emission. We have removed the bump at $1.71$ $\mu m$ for aesthetic reasons. Right panel - Zoom-in of the optical and IR HST photometry.}
    \label{fig:best_fitting_model_238}
\end{figure*}

\subsection{Comparing SED fitting to spectroscopic classification} \label{subsec:compare_to_MUSE}
Using SED fitting as a method to determine stellar parameters is not as accurate as a full spectroscopic approach. In order to benchmark the accuracy of the parameters returned from the best-fitting SEDs, we obtained archival optical spectra for two continuum sources, source 152 and source 354 in our sample. These sources are not part of this kinematic analysis, as only a few hydrogen lines are seen in emission. Nonetheless, they have been classified in an identical manner to the five sources being discussed here, and so should indicate how accurate our SED fitting approach is. Their optical spectra were obtained with MUSE by \cite{kuncarayakti2016unresolved}. Sources 185 and 251 were also observed during this program, but after inspecting their spectra, we found that all optical hydrogen lines were filled in with emission, just as their NIRSpec spectra are, making spectroscopic classification with these spectra impossible. For sources 152 and 354 however, both $H_{4-2}$ and a portion of the Paschen series are strongly detected in absorption. We performed fits of their optical spectra with the Phoenix stellar models to constrain $T_{eff}$ and $A(V)$. Table \ref{sed_vs_spec} shows the best-fitting $T_{eff}$ and $A(V)$ from the SED fitting and spectroscopic fitting. The best-fitting optical spectra for these sources are shown in appendix \ref{MUSE_spectra}.
\begin{table}
\caption{SED fitting (HST) compared to spectroscopic fitting (MUSE).}
\begin{tabular}{@{} l *4c @{}}
\toprule
 \multicolumn{1}{c}{ID}    &SED $T_{eff}$(K)& MUSE $T_{eff}$(K)& SED A(V)& MUSE A(V)\\ 
\midrule
 152&  $6185$ & $7200$ & 3.533 & $4.53$\\
 354&  $6875$ & $7000$ & 4.842 & $5.16$\\\bottomrule
 \end{tabular}
 \label{sed_vs_spec}
\end{table}
In both cases we underestimated the true $T_{eff}$ of the sources, which have optical spectra consistent with spectral type A. This appears consistent with our indirect arguments given above. Exceedingly high levels of veiling are required to remove metal absorption lines from the NIR spectra of cooler stars. So much so that the entire SED shape would become incompatible with our observations. Invoking a high $T_{eff}$ naturally explains the lack of metal absorption, and is broadly consistent with the relatively high temperatures that were returned by our SED fitting approach. From this, we conclude that sources $185$, $238$ and $823$ are Herbig Ae type stars, while sources $251$ and $469$ have spectra consistent with spectral type G and F respectively. Given the limitations of SED fitting, it is possible that we have underestimated the temperatures of these sources. It is less likely that any of these five sources are cooler than the best-fitting SED temperatures. Longer wavelength observations from JWST MIRI would greatly help in constraining the disc emission and allow for simultaneous fitting of the photospheric and disc properties. Alternatively, high spectral resolution optical spectroscopy could allow for the wings of filled in hydrogen absorption lines to be detected (such as in \cite{fairlamb2015spectroscopic}), enabling direct spectral classification for these sources.
\subsection{Assessment of uncertainties} \label{subsec:fitting_uncs}
The uncertainties returned from our MCMC exploration are statistical uncertainties based on $16th - 84th$ posterior distribution percentiles, the Bayesian equivalent to $1\;\sigma$ uncertainties. They represent the range of plausible parameter values given both our observations as well as our priors. Given that we used uninformative priors with a wide range of permitted values for each parameter, the priors do not contribute significantly to the final best-fit values, and ensure that our uncertainties are conservative, rather than unrealistically small. We feel this is reasonable given how little was known about these sources a priori. Our error bars do not reflect systematic uncertainties, such as those inherited from the interpolation of the Phoenix models (see \citet{czekala2015constructing} for a discussion on uncertainties arising from spectral model interpolation), the intrinsic accuracy of the Phoenix models themselves, or systematic uncertainties related to imperfect calibration of the NIRSpec spectra.\\ 
In \cite{RogersCTTS} we employed the same MCMC exploration, but fit both the SED and NIRSpec spectrum of sources with clear photospheric absorption features. The typical uncertainties for $T_{eff}$ and $A(V)$ in those cases was $58\;K$ and $0.2\;mag$ respectively. Comparing this to the error bars returned from the SED fitting of our continuum sources, we find typical uncertainties on $T_{eff}$ and $A(V)$ of $603\;K$ and $0.33\;mag$ respectively. The error bars assigned to $T_{eff}$ and $A(V)$ in this study reflect the greater degree of uncertainty associated with SED fitting.\\ 
$M_{*}$ and $R_{*}$ of each source were determined in \cite{RogersCTTS} by fitting evolutionary tracks to the sources on the Hertzsprung Russel diagram (HRD). Likewise, these error bars reflect only the statistical uncertainties on these parameters. $M_*$, $T_{eff}$ and $L_*$ of our sample had been estimated before in \cite{beccari2010progressive} based on a colour-magnitude diagram using the $F555W$ and $F814W$ filters. In that study, a single value of $A(V)\;=\;4.5\;mag$ was applied to all sources to de-extinguish them. The typical uncertainty on $M_{*}$ was a factor $\sqrt{2}$, corresponding to uncertainties on $T_{eff}$ of $1000\;- 2000\;K$ (private communication, De Marchi.). As we fit each source individually, using more data points and constraints than \cite{beccari2010progressive}, we believe that the best-fitting parameters returned by our MCMC exploration are the most accurate available for these sources currently.

\section{Methods - Measuring the recombination lines}
\label{sec:measure_lines}
\begin{figure}[h!]
    \centering
    \includegraphics[width=1\linewidth]{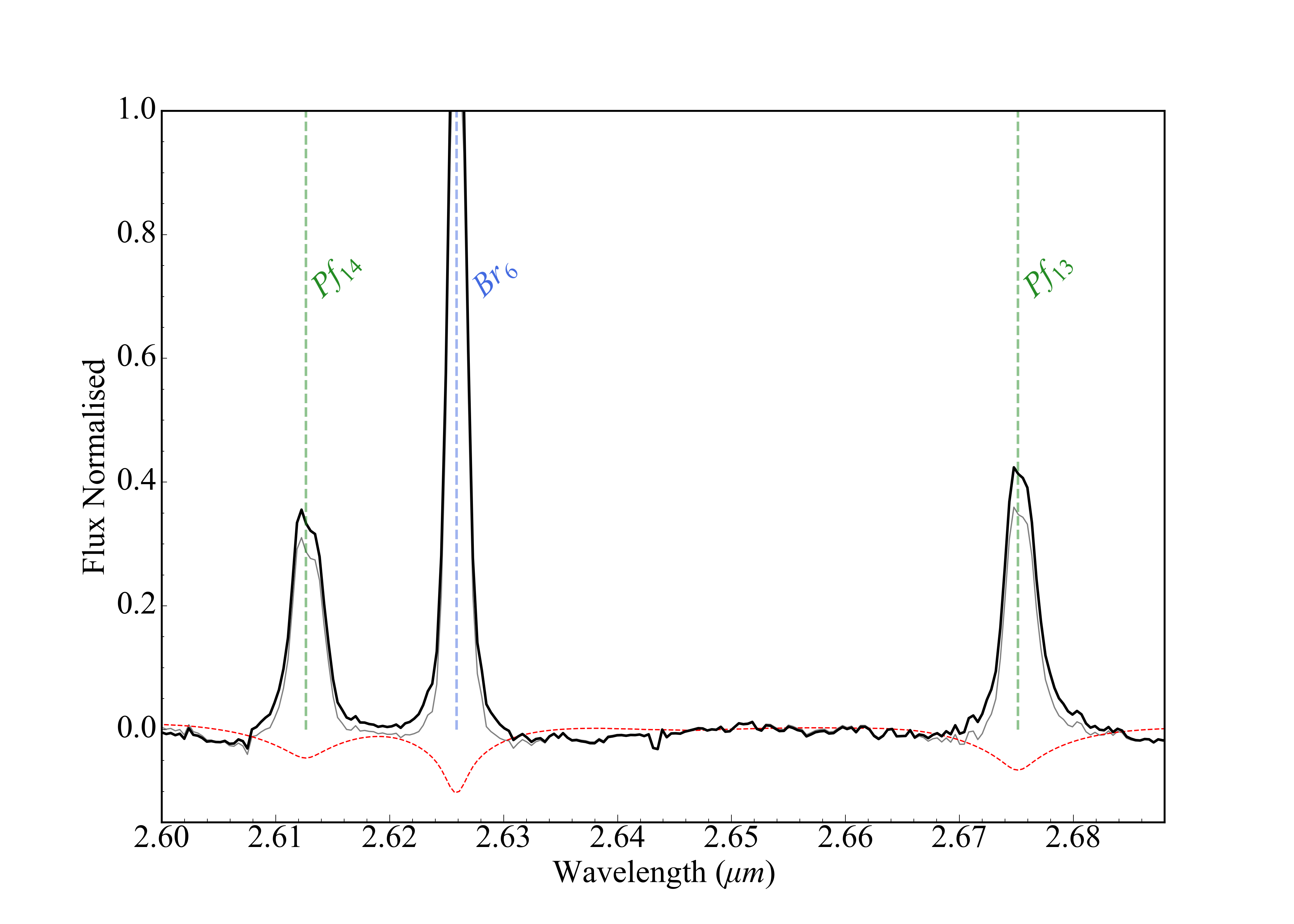}
    \caption{Section of the normalised spectrum of source 238. The grey thin line shows the observed spectrum. The red dashed line shows the photospheric model spectrum after correcting for veiling. The black thick line shows the observed spectrum after subtraction of the photospheric absorption lines.}
    \label{fig:norm_spectra}
\end{figure}

In order to measure the hydrogen emission lines in the spectra, we employed Monte Carlo simulations. This allowed for the straightforward propagation of uncertainties that arose from our processing steps and corrections. To do this, we generated $1000$ realisations of each emission line, as well as the corresponding absorption lines from the underlying stellar photosphere. The lines were allowed to vary within their uncertainties for each realisation. The absorption lines were varied by generating a new synthetic photospheric spectrum with varying $T_{eff}$, $A(V)$ and $r_{\lambda}$ based on our results in Section \ref{sec:sed_fitting}. In order to reveal the true emission line profiles, the underlying absorption lines needed to be subtracted. This was done by normalising both the observed spectrum and photospheric model spectrum, after correcting for extinction and veiling. Figure \ref{fig:norm_spectra} shows a section of the normalised spectrum of source 238 after subtraction of the photospheric absorption lines. 

We performed a fit of a Gaussian profile to each realisation of the emission lines using the non-linear least squares fitting routine curve fit from SciPy. We calculated the equivalent width (EW) and FWHM of the best-fitting Gaussian, and converted the EW to a flux by multiplying it by the adjacent continuum. The final flux and FWHM of each line is the median of the $1000$ measurements and we took as its uncertainty the standard deviation of the $1000$ measurements. Appendix \ref{emission_line_tables} lists the emission line EWs and luminosities for each source. In figure \ref{fig:Herbig_spectra}, there is a noticeable bump in the spectra at $\sim 1.71 \mu m$, coincident with the $Br_{10}$ line. This is a calibration artefact that is introduced after the filter throughput correction step is performed in the pipeline. The line profiles, uncertainties, and best-fitting Gaussian profiles for source 238 are shown in figure \ref{fig:238_profiles}. The line profiles for the remaining sources are shown in appendix \ref{line_profiles}.
\begin{figure*}[h]
    \centering
    \includegraphics[width=0.8\linewidth]{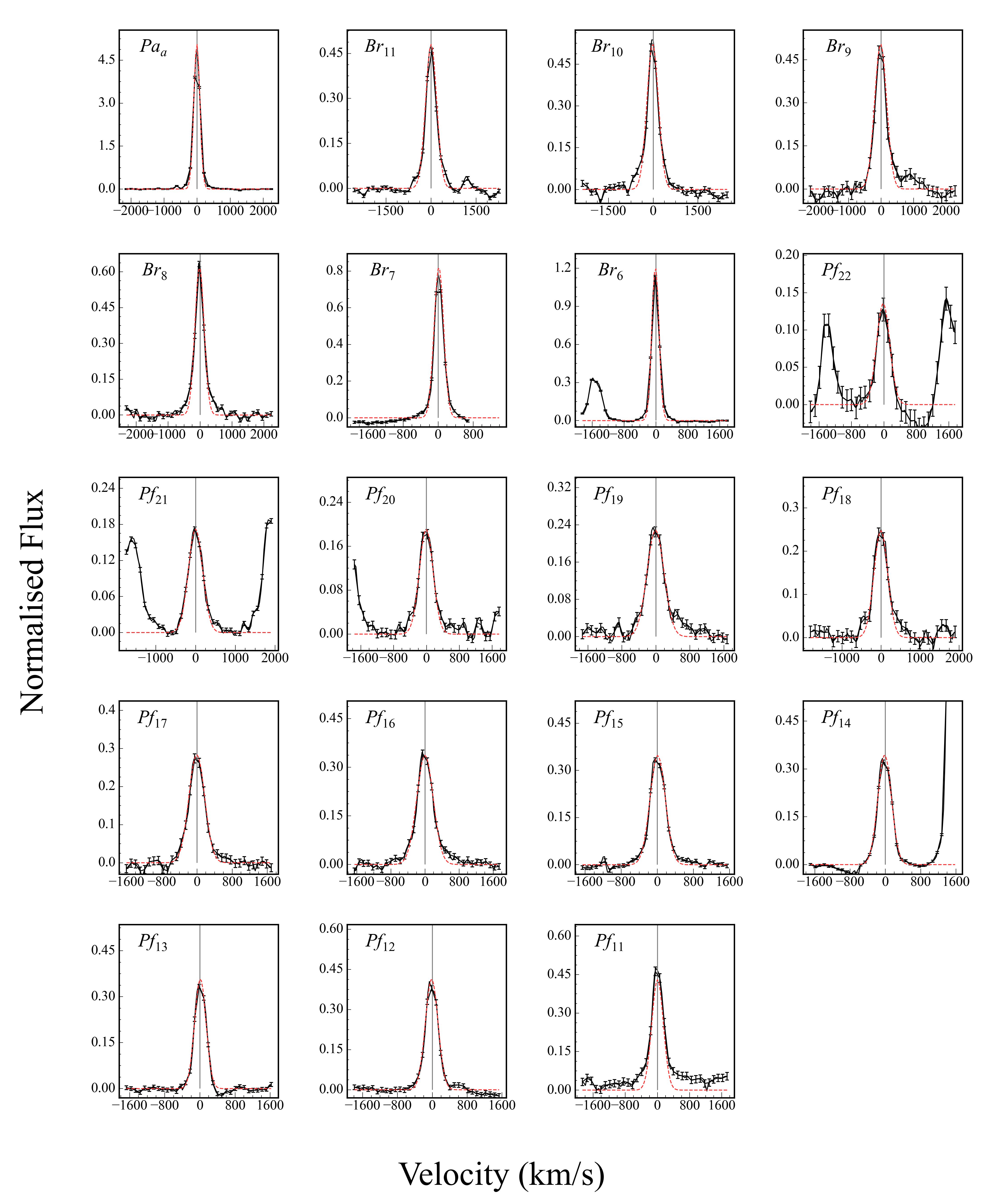}
    \caption{Fitted hydrogen line profiles for source 238. The black solid line is the normalised line profile with the continuum set to zero. The error bars represent the $1\;\sigma $ uncertainties on the flux. The red dashed line is the best-fitting Gaussian profile. The transition name is given in each plot.}
    \label{fig:238_profiles}
\end{figure*}
\section{Results - Line kinematics}
\label{sec:results}
\subsection{FWHM of hydrogen emission lines}
\label{subsec:fwhm_h_lines}
\begin{figure}
    \centering
    \includegraphics[width=1\linewidth]{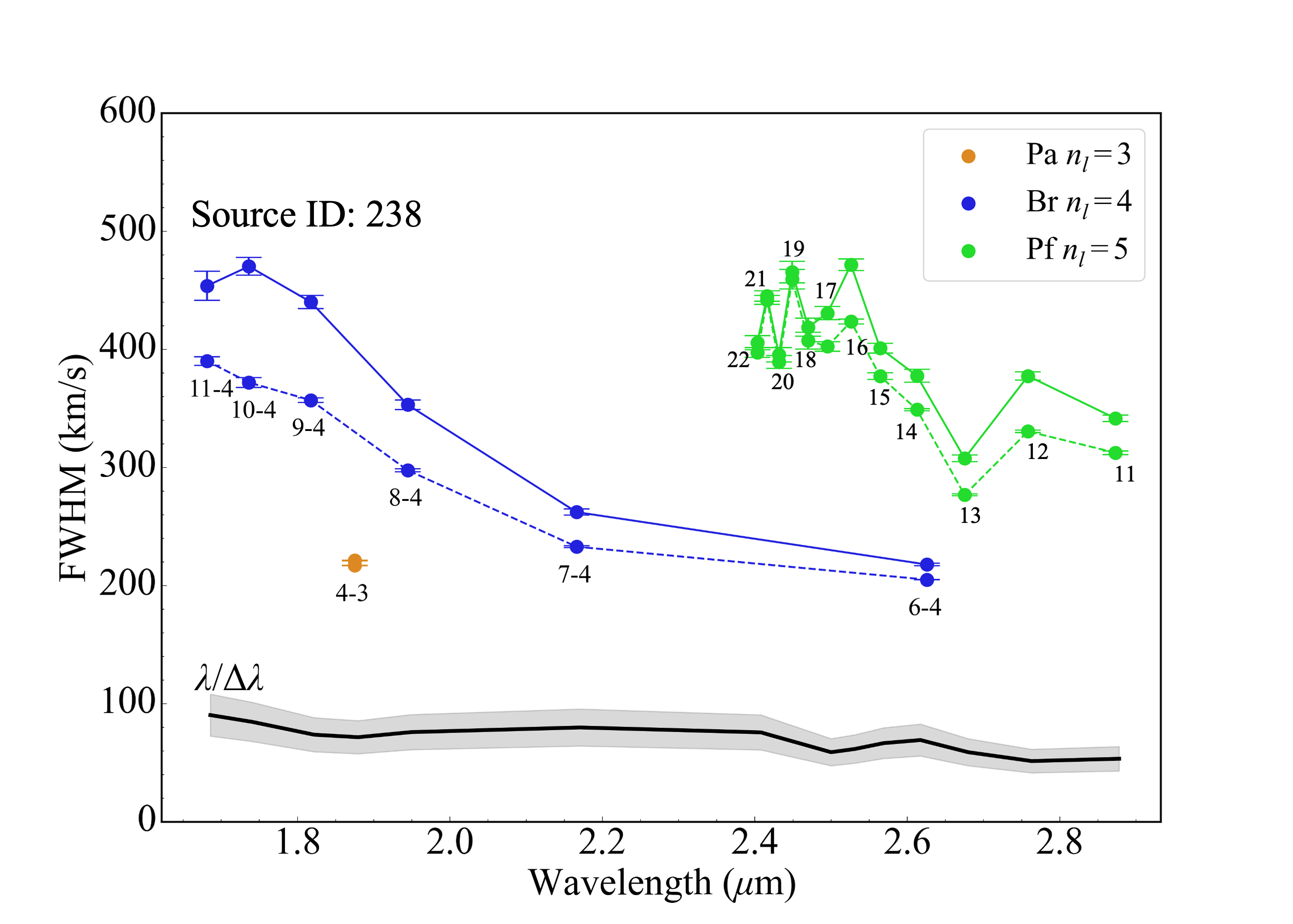}
    \caption{Full width at half maximum of each hydrogen emission line of source $238$. The Paschen series is shown in orange, and the Brackett series is shown in blue. The dashed line is before photospheric subtraction, and the solid line is after photospheric subtraction. The Pfund series is shown in green. The dashed line is before photospheric subtraction. The solid line is after photospheric subtraction. The expected instrumental resolution is shown in black. The systematic uncertainty of the spectral resolution is in  pale grey. Transition names are indicated. For the Pfund transitions, only $n_{up}$ is shown for visual clarity.}
    \label{fig:238_FWHM}
\end{figure}

Emission lines are broadened through numerous processes, including natural broadening, thermal broadening, rotational (Doppler) broadening and pressure broadening (eg. Stark broadening). In general, these processes contribute $\le 100 km s^{-1}$ of broadening (although in some cases it can be significantly more as we discuss in Section \ref{subsec:stark}). In the circumstellar environment, gas moving at high velocities due to infall and outflow processes can introduce significant line broadening, of order hundreds of $km s^{-1}$. As such, the FWHM of emission lines from PMS stars is typically interpreted as reflecting the velocity of the gas that produced the emission line. Narrow lines then originate from slow moving gas (e.g. molecular outflows of $H_2$ and $CO$ with $FWHM \le 10 km/
 s^{-1}$), while broad lines originate from high-velocity gas (e.g. H I, $FWHM \ge 100 km s^{-1}$). Figure \ref{fig:238_FWHM} shows the FWHM of the Paschen, Brackett and Pfund emission lines from source $238$. We have also plotted the resolving power of NIRSpec obtained in Section \ref{subsec:spec_res}. Although we focus on source 238 for the remainder of this analysis and discussion, the same kinematic behaviour is seen for all five sources. The FWHM diagrams of the other sources can be seen in the Appendix \ref{subsec:FWHM_diagrams}.\\ 
A striking relationship is immediately apparent. The FWHM for a given series tends to decrease with increasing wavelength. This is evidently not purely a wavelength effect, as lines from different series with similar wavelengths have drastically different FWHMs. Rather, the trend appears to follow the upper energy level $n_{up}$ of the transition, with higher order transitions (electrons falling from high $n_{up}$) tending to have larger FWHMs. This result suggests that the high excitation lines come from the fastest moving gas, while lower excitation lines come from gas with somewhat lower velocities. The Pfund series displays some significant scatter, as the highest order lines are bunched close together in wavelength and are much weaker than the Brackett or Paschen series. This makes the measurement of these lines challenging in terms of identifying the true level of the continuum, and therefore obtaining reliable fits to the weakest lines. The kinematic trend is dramatic enough that it is still clearly present despite these challenges.
\subsection{Line profile optical depths}
\begin{figure}[h!]
\begin{subfigure}{1\linewidth}
    \centering
    \includegraphics[width=1\textwidth]{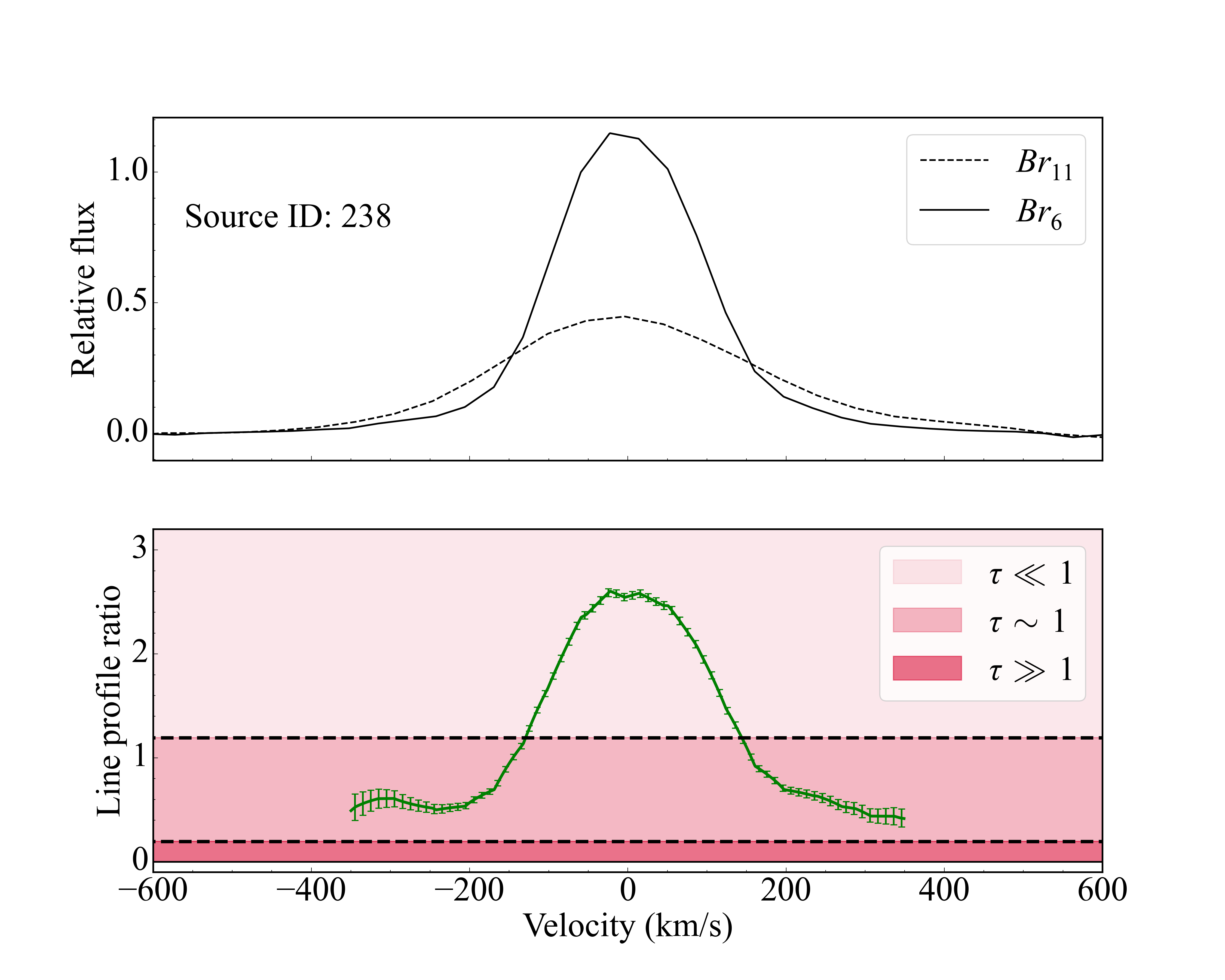}
    \label{fig:tau_Br}
\end{subfigure}
\begin{subfigure}{1\linewidth}
    \centering
    \includegraphics[width=1\textwidth]{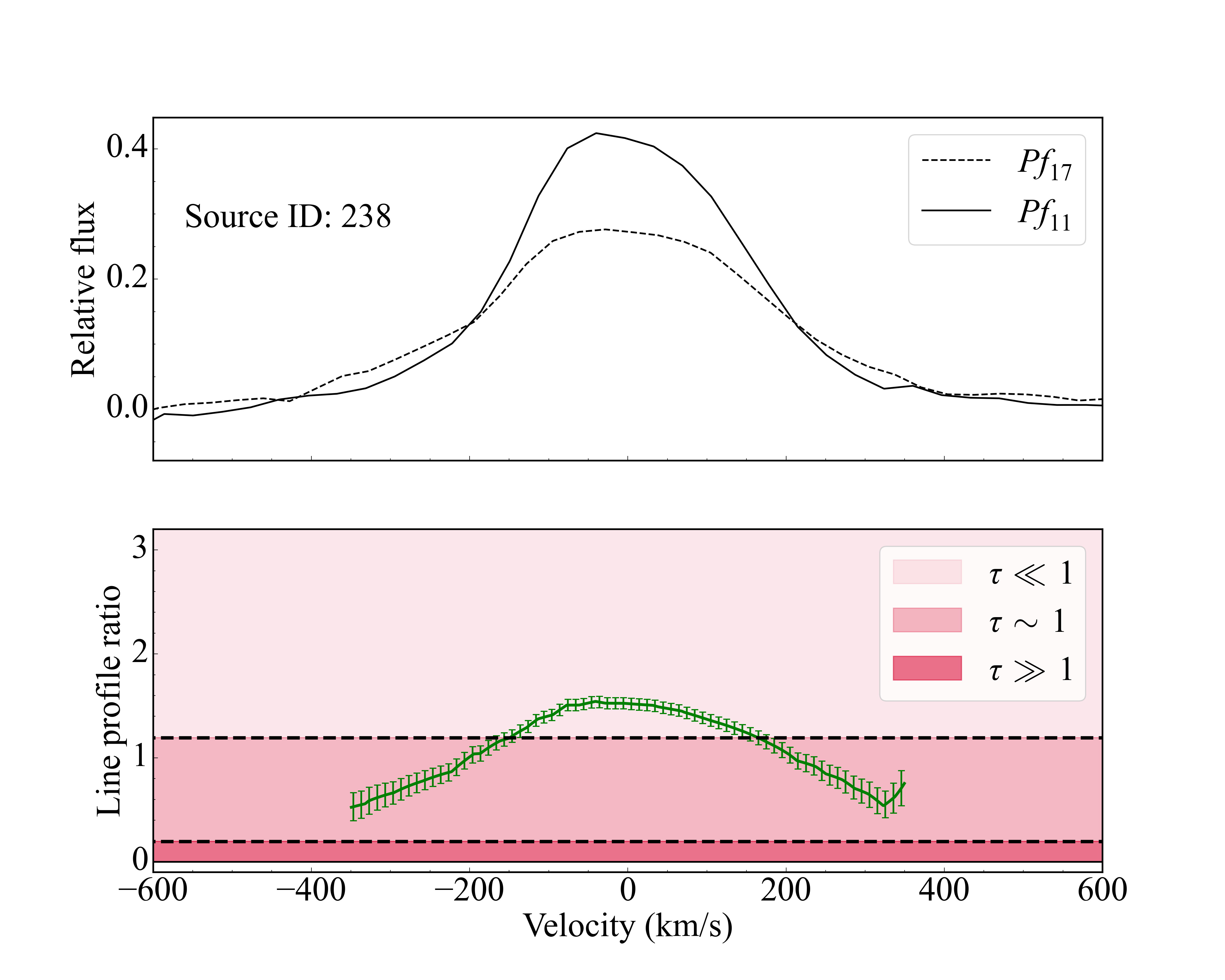}
    \label{fig:tau_Pf}
\end{subfigure}
        
\caption{Optical depth diagrams for spectrally resolved emission lines. \textbf{Top}: Line profiles $Br_{11}$ and $Br_{6}$ are shown as the dashed and solid black lines, respectively. The line profiles are plotted on a velocity scale, with each line centred at its rest wavelength. Underneath, the green solid line shows the line profile ratio $\frac{Br_{11}}{Br_{6}}$. The dark red patch represents the fully optically thick values. The medium red patch represents the partially optically thick values. The light red represents the optically thin values. \textbf{Bottom}: Same as above but for $Pf_{17}$ and $Pf_{11}$.}
\label{fig:238_optical_depth}
\end{figure}

An additional way to investigate the kinematics of the hydrogen emission lines is to take the ratio of two spectrally resolved line profiles, in order to probe the optical depth of the lines at various velocities. This analysis is akin to that performed by \cite{bunn1995observations} and \cite{lumsden2012tracers}. We have used two pairs of lines in this analysis, $\frac{Br_{11}}{Br_{6}}$ and $\frac{Pf_{17}}{Pf_{11}}$. These lines were chosen so that the optical depth of the Brackett and Pfund series could be compared to each other, as well as to compare lines with significantly different $n_{up}$/excitation energies. If the lines are optically thin, the line profile ratio should be consistent with Case B recombination \citep{baker1938physical}. If the emission is completely optically thick, then the ratio simply becomes a function of wavelength and emitting area:
\begin{equation} \label{form_opt_depth}
    \centering
    \frac{I_{1}}{I_{2}} = (\frac{\lambda_{2}}{\lambda_{1}})^4(\frac{S_{1}}{S_{2}}),
\end{equation}
where $I$ is the intensity of the emission line, $\lambda$ is the wavelength of the line, and $S$ is the emitting surface area. This equation is taken from \cite{lumsden2012tracers}, page 1095.\\
One would expect that the emitting area $S$ should be smaller for the high excitation lines. The degree of excitation in the gas increases closer to the central star. A relatively small portion of gas in the accretion flow or wind is excited enough to produce high excitation lines, while a larger, more extended portion of this gas farther away from the central star is only excited enough to produce lower excitation lines. For simplicity however, we assumed that $S_{1} = S_{2}$, which provides us with a lower limit to the optically thick ratio for each pair of lines. Line ratios that lie between the completely optically thick and optically thin regimes correspond to only one of the two lines being optically thick. For the optically thin limit for a given ratio, we have used the tables provided in \cite{storey1995recombination}. We considered the full range of electron temperatures ($500 - 30,000\;K$) and densities ($10^2-10^{12}\;cm^{-3}$), and adopted the smallest value. This represents the ratio that is closest to optically thick while still being consistent with optically thin emission. In other words, an upper limit. The optically thick limit has been computed based on equation \ref{form_opt_depth}, and as discussed, represents the lower limit.\\
Figure \ref{fig:238_optical_depth} shows the optical depth diagrams for $\frac{Br_{11}}{Br_{6}}$ and $\frac{Pf_{17}}{Pf_{11}}$ for source 238. The optical depth diagrams are shown for the remaining sources in appendix \ref{subsec:optical_depth_appendix}. The uncertainty of the line ratios are shown as green error bars, based on the uncertainties of the individual emission line fluxes. Due to the limited number of pixels sampling each line, we interpolated the line profiles so that they could be accurately centred and aligned. As a result, it is important not to over-interpret these diagrams. Small scale variations in the line profile ratio could be artificial, and likely do not reflect real physical changes in the line optical depth over small velocity scales. Only the overall shape of the line profile ratio should be considered, and where it is clearly optically thin, and clearly optically thick.
From these optical depth diagrams, we can see that for both the Brackett and Pfund lines, the lowest-velocity components of the lines are consistent with optically thin emission. The wings of the lines, however, quickly become optically thick, with the highest-velocity components of each line showing the most optically thick ratio. Interestingly the Pfund lines show a much larger fraction of the line profile ratio in the optically thick regime compared to the Brackett lines.
\section{Discussion - The origin of circumstellar hydrogen emission lines}
\label{sec:discussion}
\subsection{Magneto-centrifugal winds}
\begin{figure}[h]
    \centering
    \includegraphics[width=1\linewidth]{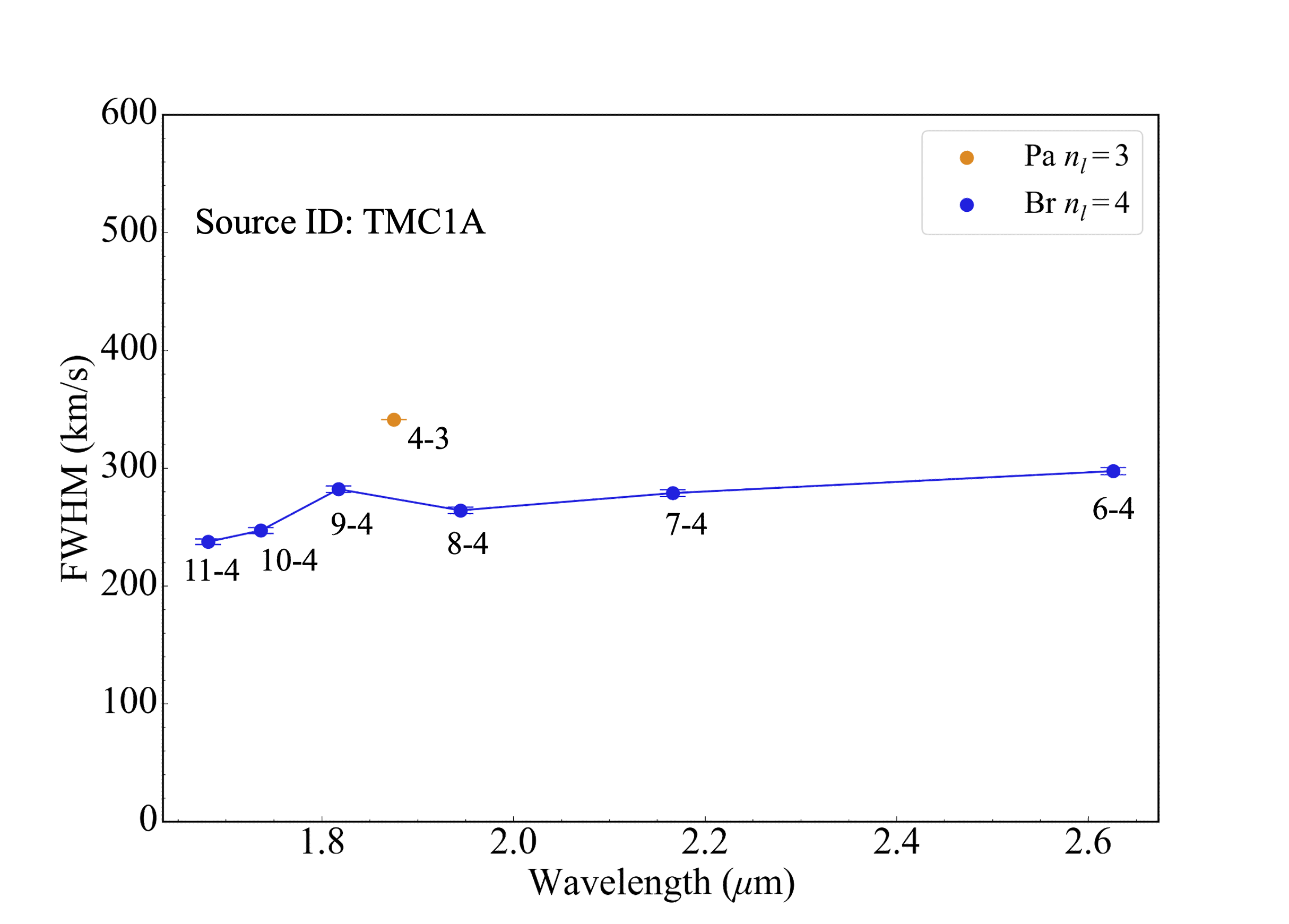}
    \caption{Full width at half maximum of each hydrogen emission line of source $TMC1A$. The Paschen series is in orange, and the Brackett series is in blue.}
    \label{fig:FWHM_TMC1A}
\end{figure}

\begin{figure}[h]
    \centering
    \includegraphics[width=1\linewidth]{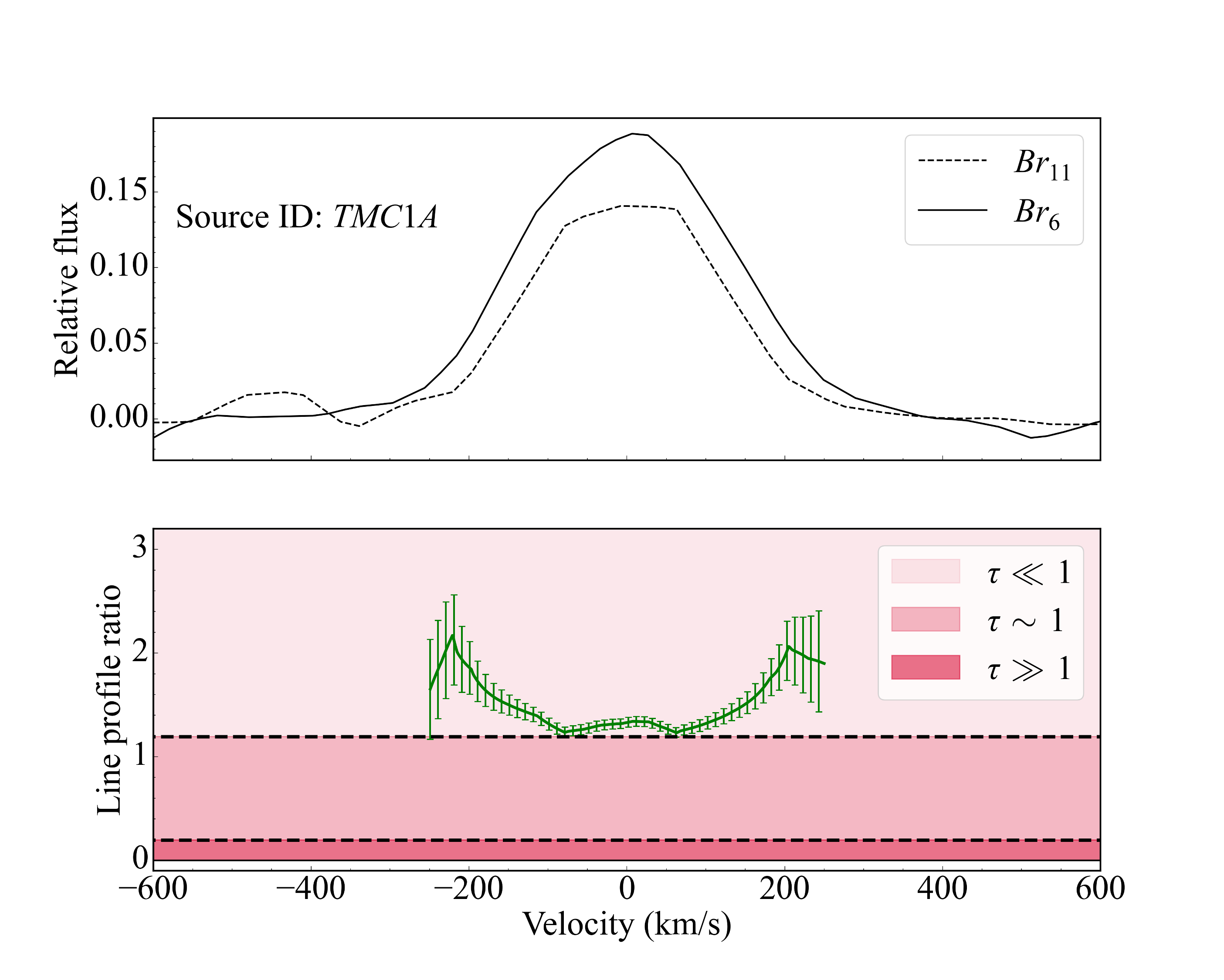}
    \caption{Line profiles of $Br_{11}$ and $Br_{6}$, respectively shown as solid and dashed black lines. The green solid line shows the line profile ratio $\frac{Br_{10}}{Br_{6}}$. Line profile ratio uncertainties are shown as green error bars.}
    \label{fig:tau_TMC1A}
\end{figure}

The two most common origins suggested for hydrogen emission lines from young stars are either the accretion flow and/or an accelerating magneto-centrifugal disc wind \citep{muzerolle1998magnetospheric, kurosawa2006formation, lima2010modeling}. If a magneto-centrifugal wind were the dominant line emission mechanism, the high excitation lines should come preferentially from close to the launching point near the disc surface, where the gas is hottest and most strongly irradiated by the central star. The gas at this point should then have a low velocity, as it has not yet experienced much acceleration. Likewise, one would expect the lower excitation lines to come preferentially from the gas at greater distances from the star, as the irradiation/temperature is still sufficient to produce low excitation lines, and the effective emitting area is significantly larger than at the disc surface. This gas would then have a high velocity, as it has experienced more acceleration. In this picture, the high excitation lines should display narrow profiles, and the low excitation lines should display broad profiles.\\
This general trend has been observed in \cite{nisini2004observations} who observed a very young Class I protostar, known to exhibit powerful jets and outflows, with the lower excitation $Br_{7}$ line appearing significantly broader than the higher excitation lines $Br_{12}$ and $Br_{13}$. We have analysed archival JWST NIRSpec IFU data of source $TMC1A$, a class I protostar, with a spatially resolved outflow (see \citet{harsono2023jwst} for a full discussion of this source). These observations included the same grating and filter combination as our own, allowing for a direct comparison to our observations. Given that TMC1A is spatially resolved, we expect that the resolving power of these observations are intermediate between our own and that of a uniformly illuminated aperture. We have performed the same kinematic analysis on these data, extracting the flux from the datacube using a $3 \times 3$ pixel aperture. We have made use of only the Paschen and Brackett lines. The Pfund lines in this case are either too weak or located in the strong $H_20$ absorption feature at $3.08 \; \mu m$, and therefore they could not be measured reliably. Figure \ref{fig:FWHM_TMC1A} shows the resulting FWHMs for each line.
The trend seen here is clear. The highest excitation line $Br_{11}$ has the smallest FWHM, $\sim 240 km s^{-1}$, while the lowest excitation line $Pa_{\alpha}$ has the highest FWHM, $\sim 340 km s^{-1}$.\\
In an accelerating wind, the highest-velocity gas far away from the star should have low optical depth, as the density has fallen off by several orders of magnitude compared to the star/disc surface \citep{lima2010modeling}. Conversely, the gas at the launching point of the wind, with much higher density, should have higher optical depth. The resulting line profile ratios would then show higher optical depth in the line core, becoming optically thinner towards the wings. \\
In figure \ref{fig:tau_TMC1A}, we show the line profile ratio for $Br_{11}$ and $Br_{6}$ for TMC1A. The optical depth diagram of TMC1A is essentially inverted compared to our sources. The line core is the closest to optically thick, while the wings are firmly in the optically thin regime. We note that we have not attempted to correct the spectrum of TCM1A for extinction. Doing so would cause the flux of $Br_{11}$ to increase by a greater amount than $Br_{6}$, pushing the core of the line profile ratio further into the optically thick regime.\\ 
The kinematic trends seen here suggest that the hydrogen emission lines in TMC1A come predominantly from an accelerating wind, which is supported by the spatially resolved outflow seen towards this source, as well as the presence of other wind and outflow tracers such as $H_2$ and $Fe [II]$ emission. This conclusion was also reached by the authors \citep{harsono2023jwst}.
\subsection{Magnetospheric accretion} \label{subsec:magneto_acc}
The trends seen for TMC1A in terms of kinematics and optical depth are in total contrast to what we observed towards our sources in NGC 3603. Thus, we now consider the accretion scenario as the dominant line emission mechanism for the sources in NGC 3603. In this scenario, the high excitation lines should come preferentially from gas in the accretion flow close to the stellar surface. Here the conditions are preferable for high excitation line emission, as the gas is strongly irradiated by the star and the temperature is significantly higher than at the disc surface \citep{lima2010modeling}. This gas would move with high velocity, as it has been in near free-fall from the disc surface and hence has experienced significant acceleration. The low excitation lines should come preferentially from gas at the base of the accretion flow on the disc surface. Here the temperature and irradiation are still sufficient to produce low excitation emission lines, and due to the funnel-like geometry of the accretion flow the effective emitting area is larger at the base of the flow compared to just before it hits the stellar surface \citep[e.g.][]{muzerolle1998magnetospheric}. This gas would move with a lower velocity as it has yet to experience much acceleration. In this picture, the high excitation lines should display broad profiles, as they come from the highest-velocity gas, while the low excitation lines should display narrow profiles, as they come from the lower velocity gas. This is consistent with the emission line profiles that we have measured across three spectral series, for over $15$ hydrogen emission lines for source 238, as well as for the other four sources in this sample.\\
The fastest moving gas in an accretion flow should also be the most optically thick, as the density of the gas in the accretion flow increases as it approaches the central star. Again, this is consistent with the line profile ratios measured for our sources.\\ 
To further test the accretion scenario, we have calculated what the free-fall velocity ($v_{ff}$) would be for each source. $v_{ff}$ is given by the equation

\begin{equation} \label{v_ff}
    \centering
    v_{ff} = \sqrt{2G\;M_{*}(\frac{1}{R_{*}} - \frac{1}{R_{trunc}})} \; cos(i)\;sin(\theta),
\end{equation}

where $i$ is the inclination angle and $\theta$ is the angle of the accretion material with respect to the equatorial plane. For all calculations we fixed it at $\theta=\frac{\pi}{2}$, corresponding to material reaching the stellar surface at the poles of the star. $R_{trunc}$ is the truncation radius, the radius at which the magnetosphere reaches into the disc. $R_{trunc}$ is often assumed to be equal to $5R_{*}$ for CTTS \citep{hartmann2016accretion}, but is likely smaller for Herbig AeBe stars, a result of their weaker magnetic fields \citep{mendigutia2020mass}. We have assumed a truncation radius of $3\;\pm \;1R_{*}$ to account for this. $v_{ff}$ is determined by $M_{*}$ and $R_{*}$, which we have determined via SED fitting of our sources. Given the higher degree of uncertainty associated with SED fitting, as well as the systematic uncertainties associated with stellar evolutionary models, we have taken the conservative approach of  employing $3\sigma$ uncertainties for $M_{*}$ and $R_{*}$ in order to determine uncertainties on $v_{ff}$.\\ 
Table \ref{free_fall_velos} lists $v_{ff}$ of each source when the gas makes contact with the stellar surface. We have considered three inclination angles of $0^{\circ},\; 30^{\circ},$ and $60^{\circ}$. An observer would measure a maximum line of sight velocity from the accretion flow when viewing the system face-on, with $i\;=\;0^{\circ}$. We also list the FWHM of the $Br_{11}$ line, which represents a line from some of the highest-velocity gas that we could measure with low statistical uncertainty.\\
The maximum $v_{ff}$ are consistent with the FWHM of $Br_{11}$ for all sources. This demonstrates that gas in a MA flow in free-fall is capable of producing the observed line widths for our sample. This result, in combination with the kinematic trends that we have presented strongly point towards MA over magneto-centrifugal winds. However, other line emission mechanisms can be present around young stars, and so these must be addressed before we can claim that MA is the dominant line emission mechanism for our sources.

\setlength{\tabcolsep}{2pt} 
\begin{table}
\caption{Free-fall velocities for each source for different inclination angles.}
\centering
\begin{tabular}{@{} l *4c @{}}
\toprule
 \multicolumn{1}{c}{ID} &$v_{ff}\;(km/s)\;0^{\circ}$& $v_{ff}\;(km/s)\;30^{\circ}$ & $v_{ff}\;(km/s)\;60^{\circ}$ & FWHM $Br_{11}$ \\ 
\midrule
 185&  $335\;\pm \; 128$ & $290\;\pm\;110$ & $167\;\pm \;64$ & $281\; \pm\; 18$\\
 238& $382\;\pm \; 108$  & $331\;\pm\;94$  & $191\;\pm \;54$ & $452\; \pm\; 17$\\
 251&  $301\;\pm \; 195$ & $261\;\pm\;169$ & $150\;\pm \;98$ & $206\; \pm\; 3$ \\
 469&  $329\;\pm \; 115$ & $284\;\pm\;100$ & $164\;\pm \;58$ & $220\; \pm\; 2$ \\
 823& $385\;\pm \; 76$   & $334\;\pm\;66$  & $192\;\pm \;38$ & $373\; \pm\; 5$ \\ \bottomrule
 \end{tabular}
 \tablefoot{\small{The uncertainties given here are based on $3\;\sigma$ uncertainties on $M_{*}$ and $R_{*}$ and the range of truncation radii assumed.}}
 \label{free_fall_velos}
\end{table}
\subsection{Alternative line emission mechanisms}
We have focused on accretion and magneto-centrifugal star/disc winds to try and explain the hydrogen emission lines observed towards our Herbig AeBe stars. There are, however, alternative mechanisms that could potentially explain the kinematic and optical depth results that have been presented so far. They include
(1) line driven stellar winds \citep{kudritzki2000winds}, (2) internal/external photoevaporative winds \citep{williams2011protoplanetary, winter2022external}, (3) a hot gaseous inner disc \citep{muzerolle2004magnetospheres}, and (4) boundary layer accretion \citep{lynden1974evolution}.

\subsubsection{Line-driven stellar wind}
Line driven stellar winds are common around massive stars, whose stellar luminosity exceeds the Eddington limit \citep{abbott1982theory}. Here stellar photons impart enough momentum to gas on the stellar surface to overcome the gravitational potential of the star, driving a decelerating wind. A decelerating wind could qualitatively match the kinematic results presented. However, although our stars are more massive and hotter than CTTS, their luminosities are still well below the Eddington limit, and hence line driven winds are not expected around our sources. \\
\subsubsection{Internal photoevaporative wind}
X-ray and far-ultraviolet radiation from the central star can drive thermal, decelerating winds from the disc surface \citep{owen2012theory}. These winds can produce hydrogen emission lines that could in principle display the same kinematic trends that we have measured. \cite{ercolano2010theoretical} computed theoretical line luminosities for many hydrogen recombination lines forming in a photoevaporative wind. For $Br_{7}$, the line luminosity $\frac{L(Br_{7})}{L_\odot}$ ranges from $3.05 \times 10^{-9}$ erg s$^{-1}$ to $1.26 \times 10^{-7}$ erg s$^{-1}$, for X-ray luminosities $L_X$ of $10^{29.3}$ erg s$^{-1}$ to $10^{30.3}$ erg s$^{-1}$, respectively. These calculations were carried out for CTTS. X-ray luminosities are generally higher for Herbig AeBe stars compared to CTTS, with  $L_X$ up to $\sim 10^{32}$ erg s$^{-1}$ \citep{zinnecker1994x, hamaguchi2005x}. This would likely lead to larger internal photoevaporative $Br_{7}$ line luminosities. However, these values are still $4-6$ orders of magnitude lower than the line luminosities measured for $Br_{7}$ in our sample. As such, we do not expect that internal photoevaporative winds could dominate the hydrogen line emission for our sources.\\ 
\subsubsection{External photoevaporative wind}
Since our sources are located in a massive star forming region, external photoevaporation could contribute towards the hydrogen line emission. The massive stars of spectral type O and B  at the centre of NGC 3603 irradiate our sources, which can result in a thermally driven cocoon of gas and dust escaping the protoplanetary disc, with an ionisation front around the source resulting from extreme-ultraviolet photons. This cocoon morphology is seen commonly around the PMS stars in Orion known as proplyds \citep[e.g.][]{ricci2008hubble}. At a distance of $7.2$ kpc, it is not possible to spatially resolve these sources with current facilities, and so it is not obvious whether they possess this cocoon morphology or ionisation front. However, in the case of Orion, where the detection of these features is unambiguous, the hydrogen emission lines from the ionisation front resemble the nebular emission lines, with narrow line profiles (FWHM $\le 50$ km s$^{-1}$). Furthermore, our nebular subtraction has removed any contribution from the ionisation front thanks to the He I scaling method, as the He I emission line strength would have a contribution from both the extended nebular gas as well as the ionisation front. By removing the He I emission, we have removed the contribution of both the nebular emission and any possible ionisation front emission. Based on the broad lines that we measure from our sources, and the subtraction method that we have developed and employed, an externally driven photoevaporative wind cannot be the dominant emission line mechanism in this case.\\

\subsubsection{Hot inner gaseous disc}
\begin{figure}[h]
    \centering
    \includegraphics[width=1\linewidth]{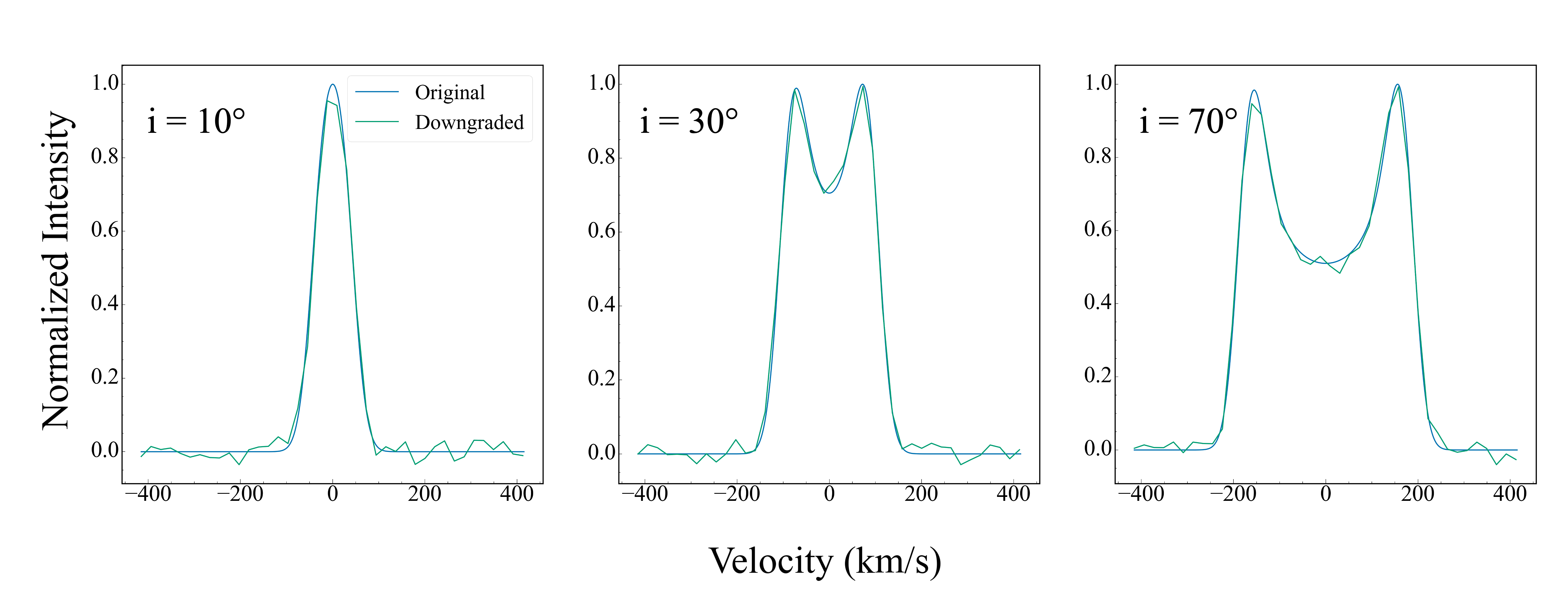}
    \caption{Line profile of $Br_{7}$ if it were to originate in a Keplerian rotating inner disc. Inclination angles are $10^{\circ}$, $30^{\circ}$, and $70^{\circ}$.}
    \label{fig:rotation_profile}
\end{figure}

The hot inner gaseous disc may contribute towards the observed hydrogen emission lines. In their seminal work \cite{muzerolle2004magnetospheres} showed that the inner disc is expected to become an important contributor of emission for mass accretion rates of $\ge 10^{-7} M_\odot $yr$^{-1}$. In \cite{RogersCTTS} we determined $\dot{M}_{acc}$ for all of the sources, including the five presented here, finding accretion rates of $1.6-6\times10^{-5}$ $M_{\odot} yr^{-1}$, well above the level indicated in \cite{muzerolle2004magnetospheres}. The inner disc as an important source of emission is also supported by the presence of $CO$ bandhead emission detected towards three of the five sources (see figure \ref{fig:Herbig_spectra}), which likely arises from the warm inner disc \citep[e.g.][]{ilee2013co}. The detection of $CO$ bandhead emission (along with the high values of $\dot{M}_{acc}$) suggests that there is still ample gas in the inner disc around the stars. In \cite{mendigutia2017compact}, spectro-interferometric observations of two Herbig AeBe stars revealed spatially unresolved $H_{\alpha}$ emission, consistent with an origin from a compact inner disc. For one of the two sources, the observations could also be reproduced with a MA model.\\ 
If the hydrogen lines were formed in the inner disc, rotational broadening would likely be the dominant line broadening mechanism, since it is the only mechanism capable of producing the FWHMs that we have measured. However, rotational broadening is often associated with a characteristic double-peaked profile in emission lines. Even with the moderate spectral resolution of NIRSpec, double-peaked profiles should be detected, even at low inclinations. \\
We created a simple toy model of $Br_{7}$ emission from the inner disc to demonstrate this. Taking the stellar parameters for source 238 from Section \ref{sec:sed_fitting}, we modelled a disc in Keplerian rotation around a star with $M_{*} = 7 M_{\odot}$ and $R_{*} = 12 R_{\odot}$. The disc inner radius $R_{inner}$ was set to $2 R_{*}$, and the outer radius $R_{outer}$ was set to $1 AU$, equivalent to $18 R_{*}$, with 100 steps between the two. The Keplerian velocity field was used to calculate the Doppler shift experienced by the line for each step in radius. We modelled the $Br_{7}$ emission line as an unresolved Gaussian (FWHM = 50 km s$^{-1}$). A Doppler shift was applied to the Gaussian for each step in radius and  the resulting Gaussians were summed together, producing the final emission line profile, normalised to its peak intensity. The resulting line profile was double peaked. We then downgraded the double peaked profile to the spectral resolution of NIRSpec at the wavelength of $Br_{7}$ and also injected Gaussian noise into the line, at the level of $2 \%$ (this is the noise level measured in the actual spectrum of source 238). The noisier, lower resolution line profile still exhibits clear double peaked emission line profiles down to inclinations of $20 \deg$. Below this inclination, the double peaked profile disappears and the resulting FWHM reduces to $\sim 100$ km s$^{-1}$. In figure \ref{fig:rotation_profile}, the profiles of $Br_{7}$ are shown at inclination angles of $10^{\circ}$, $30^{\circ}$, and $70^{\circ}$, respectively.
A similar result was also found by \cite{wilson2022hydrogen}, who computed rotationally broadened emission line profiles for $Br_{7}$ for different rates of rotation and inclinations. They showed that for rotational velocities of $\sim 130$ km s$^{-1}$, which is typical for Herbig AeBe stars \citep{bohm1995rotation}, the peak-to-peak separation of double-peaked $Br_{7}$ is $\sim 100$ km s$^{-1}$ for a viewing inclination of $20^\circ$, which would be resolved at NIRSpec's resolution. We note, however, that single peaked emission lines from the Herbig AeBe star HD 100546 have been attributed to emission from a Keplerian disc based on high spectral/spatial resolution observations \citep{mendigutia2015high}, for an inclination angle of $i\;=\;30^{\circ}$. The $Br_{7}$ line from that study showed a FWHM of $\sim 200\;km/s$, similar to the values in our sample. To further test whether the line emission from our sources could find its origin directly from the Keplerian disc, we have calculated the velocity ($v_{kep}$) of gas in Keplerian orbit around each source, similarly to our approach in Section \ref{subsec:magneto_acc}. We have compared $v_{kep}$ with the FWHM of $Br_{11}$ to test whether the velocities are consistent with the line widths measured for each source. $v_{kep}$ is defined as   
\begin{equation} \label{v_kep}
    \centering
    v_{kep} = \sqrt{G\;M_{*}/R_{*}} \; sin(i)\;cos(\theta),
\end{equation}
where $G$ is the gravitational constant, $i$ is the inclination, and $\theta$ is the azimuthal angle of the gas in orbit. Here, $\theta\;=\;0^{\circ}$ corresponds to gas moving towards the observer, and $\theta\;=\;180^{\circ}$ corresponds to gas moving away the observer. We again considered three values for $i$ of $30^{\circ},\; 60^{\circ},$ and $90^{\circ}$, where $i\;=\;0^{\circ}$ corresponds to a face-on disc with no line of sight radial velocity shift and where $i\;=\;90^{\circ}$ corresponds to an edge on disc with a maximum line of sight radial velocity shift. We have again employed $3\sigma$ uncertainties on $M_{*}$ and $R_{*}$ to determine the uncertainties on $v_{kep}$. Table \ref{keplerian_velos} shows the maximum $v_{kep}$ ($\theta\;=\;0^{\circ}$) for different inclinations.
\begin{table}[H]
\caption{Keplerian velocities for each source for different inclination angles.}
\centering
\begin{tabular}{@{} l *4c @{}}
\toprule
 \multicolumn{1}{c}{ID} &$v_{kep}\;(km/s)\;30^{\circ}$& $v_{kep}\;(km/s)\;60^{\circ}$ & $v_{kep}\;(km/s)\;90^{\circ}$ & FWHM $Br_{11}$ \\ 
\midrule
 185&  $92\;\pm \;34$ &  $159\;\pm\;59$ & $184\;\pm \; 68$ & $281\; \pm\; 18$ \\
 238&  $105\;\pm \;28$ &  $181\;\pm\;49$ & $210\;\pm \; 57$ & $452\; \pm\; 17$ \\
 251&  $82\;\pm \;53$ &  $143\;\pm\;92$ & $165\;\pm \; 106$ & $206\; \pm\; 3$ \\
 469&  $90\;\pm \;31$ &  $156\;\pm\;53$ & $180\;\pm \; 61$ & $220\; \pm\; 2$ \\
 823&  $105\;\pm \;19$ &  $183\;\pm\;33$ & $211\;\pm \; 38$ & $373\; \pm\; 5$ \\ \bottomrule
 \end{tabular}
  \tablefoot{\small{The uncertainties given here are again based on $3\;\sigma$ uncertainties on $M_{*}$ and $R_{*}$.}}
 \label{keplerian_velos}
\end{table}
Sources 185, 238, and 823, the three objects that we have classified as Herbig Ae stars, show FWHMs that are too large to be consistent with emission from the Keplerian disc,  even with our conservative approach of employing $3\;\sigma$ uncertainties for the stellar parameters, for a perfectly edge on geometry of $i\;=\;90^{\circ}$, and assuming the disc reaches all the way to the stellar surface. Sources 251 and 469 on the other hand are consistent with emission from both the Keplerian disc and MA.\\

\subsubsection{Boundary layer accretion}
An alternative accretion mechanism that has been invoked for Herbig AeBe stars is boundary layer accretion. In this scenario, the magnetic field from the star is either entirely absent or is too weak to truncate the protoplanetary disc. The inner disc connects to the stellar surface via the so-called boundary layer. Disc material must reduce its Keplerian velocity to match the stellar rotational velocity. This reduction in velocity, and hence kinetic energy, is converted into radiation producing the observed accretion luminosity \citep{mendigutia2020mass, wichittanakom2020accretion}. Given that material must decelerate in order to be accreted, with the highest excitation gas then moving at the slowest velocity, the observed kinematic trends cannot be explained if the bulk of emission comes from the boundary layer. However, in the boundary layer accretion scenario, a Keplerian disc extends close to the stellar surface. It is possible that the bulk of the line emission may come from the disc itself due to its larger emitting surface compared to the boundary layer. As such, even if boundary layer accretion were taking place for our sources, the kinematic information retrieved from the hydrogen emission lines may still be largely governed by the Keplerian disc. Because of this, we cannot rule out boundary layer accretion as a possibility for sources 251 and 469.
\subsection{Stark broadening} \label{subsec:stark}
Up to now, we have assumed that the dominant line broadening mechanism is Doppler broadening, due to the high-velocity bulk motion of the gas. This is often invoked to explain the line widths for CTTS and Herbig AeBe stars, as the velocities from infalling and outflowing material match the line widths well \citep[e.g.][]{muzerolle2004magnetospheres}. In some cases, line widths have been measured that are too broad to be consistent with Doppler broadening. One such example comes from  \cite{muzerolle1998magnetospheric}, who suggest that Stark broadening could explain the high-velocity wings observed towards $H_{\alpha}$, extending out as far as $\pm 500$ km s$^{-1}$. A feature of Stark broadening that is of particular importance in the context of our study is that transitions with high $n_{up}$ are more susceptible to Stark broadening, due to the weaker Coulomb force between those electrons with the nucleus. As a result, Stark broadening can induce qualitatively similar FWHM trends as those that we have presented. An example of this can be found in \cite{feldman1977emission}, who show the Stark effect on high order Balmer lines from the solar corona. Stark broadening is an important line broadening mechanism in Herbig AeBe photospheres due to their high temperatures and relatively high densities, both of which serve to amplify the Stark effect \citep[e.g.][]{popovic2001stark, dimitrijevic2007stark}. In Section \ref{sec:measure_lines}, we subtracted the Stark broadened absorption lines from the circumstellar emission lines. We also measured the FWHM of the emission lines before subtraction, in order to understand the impact of the underlying absorption lines. In general, the subtraction led to an increase in the FWHM of the emission lines, but did not change the overall trend seen in figure \ref{fig:238_FWHM}. This assured that the FWHM trends of the emission lines were not strongly influenced by the absorption component. However, given the potential degeneracy between Stark broadening and doppler broadening in the MA flow, it was prudent to understand whether Stark broadening may be an important broadening mechanism in the circumstellar environment. In \cite{bunn1995observations} and \cite{lumsden2012tracers}, the authors argue that Stark broadening may be important for Brackett lines in the circumstellar environment of intermediate and high mass stars, due to the presence of high-velocity wings reaching out to $\sim \: 500 km \: s^{-1}$. None of the emission lines in our sample exhibit such broad wings, with the broadest lines having wings out to $\sim \: 300 km \: s^{-1}$. In \cite{muzerolle1998magnetospheric}, the authors found that Stark broadening can impact $H_{\alpha}$ significantly for high densities $n_{H} \ge 10^{12} cm^{-3}$ and temperatures $T \sim 10000$K, but the effect dropped off quickly for higher order lines, with $H_{\beta}$ experiencing $50\%$ less broadening compared to $H_{\alpha}$. The same conclusion was found by \cite{tambovtseva2014hydrogen}, who modelled the $Br_{7}$ emission line as well as $H_{\alpha}$. They reasoned that the high optical depth of $H_{\alpha}$ leads to significant Stark broadening, while the lower optical depths of Brackett emission lines results in negligible Stark broadening compared to doppler broadening. Finally, \cite{wilson2022hydrogen} computed Stark broadened emission line profiles for $H_{\alpha}$, $Pa_{5}$, $Pa_{6}$ and $Br_{7}$ and found again that while the effect was significant for $H_{\alpha}$, the NIR lines were negligibly affected. Based on these previous results, as well as the lack of very high-velocity line wings in our observations, we do not expect that Stark broadening contributes significantly to our line widths.\\
We conclude that the line emission from sources 185, 238 and 823 is most consistent with MA. The emission lines for sources 251 and 469 are consistent with both MA and inner disc emission.
\subsection{The MA paradigm in Herbig AeBe stars}
The observed Paschen, Brackett, and Pfund lines originating from the MA flow for sources 185, 238 and 823 is a somewhat unexpected result. It is surprising that disc wind signatures are not consistent with our kinematic analysis, despite their expected importance for Herbig AeBe stars.\\
\cite{tambovtseva2016brackett} performed non-local thermodynamic equilibrium modelling to investigate whether accretion or winds were the dominant emission mechanism for $Br_{7}$ from Herbig AeBe stars, concluding that disc winds do indeed dominate. This was supported by a subsequent spectro-interferometric observational study of MWC 120 by \cite{kreplin2018brgamma}, who concluded again that $Br_{7}$ is formed predominantly in an extended wind around the star, not the compact magnetosphere. \cite{kraus2008origin} employed spectro-interferometry to try and spatially resolve the emission region of $Br_{7}$. They found that of the five sources observed, only one, source HD98922, displayed compact $Br_{7}$ emission, consistent with an origin from a MA flow. It is therefore rather surprising and puzzling that our kinematic results are consistent with MA. One perspective that we have not considered yet is that our sources are all found in a massive star forming region, residing near a population of more than $70$ O stars \citep{moffat1994ngc, drissen1995dense}. This type of environment is significantly different to where the most well studied, nearby Herbig AeBe stars reside. The harsh UV interstellar radiation field may influence the stars' ability to drive winds/outflows.\\ 
In \cite{reipurth1998protostellar}, the authors observed four Herbig-Haro (HH) objects close to the Horsehead nebula in Orion. All four of these stars lie in the vicinity of the O9.5 star $\sigma$ Orionis. In this study the authors found that the typically molecular outflows associated with HH objects are in fact ionised, which they reasoned is a result of external ionisation from $\sigma$ Orionis. They suggested that jet launching may be entirely quenched on the side that faces the OB cluster, as the four sources in their study show significantly less developed jet lobes on the directly irradiated side, while the more prominent lobes face away, possibly shadowed by the protoplanetary disc itself. Additionally, studies towards the Carina nebula also found evidence of externally irradiated jets \citep{reiter2013hst, reiter2016fe, reiter2022deep} due to the presence of nearby OB stars. Very recent spectroscopic follow up observations have revealed that outflows exposed to external UV irradiation become dissociated, ionised, and eventually totally ablated \citep{reiter2024illuminating}. The massive star content and UV radiation field of NGC 3603 is several orders of magnitude more extreme compared to Orion, and comparable to the Carina nebula. We speculate that the incompatibility between the hydrogen line kinematics and a magneto-centrifugal wind may be a result of the extreme environment that our sources find themselves in. The JWST NIRCam imaging could help answer this question. Outflows with projected sizes of $>400\;AU$ would be spatially resolved, allowing for an assessment of the impact of the external UV field on the formation, morphology and lifetime of jets and outflows in NGC 3603.\\
The origin of hydrogen emission lines from Herbig AeBe stars is not definitively known, and likely depends on multiple parameters including the age, mass, the accretion rate of the central source, and possibly even their external environment. We are not aware of any other studies in the literature that have investigated the kinematic properties of hydrogen emission lines systematically as we have presented here, and suggest that this approach could be helpful in distinguishing between different emission line mechanisms. \\ 
We also wish to emphasise that this kinematic analysis does not rely on having JWST NIRSpec observations. Although NIRSpec is an ideal facility for this type of study given its unbroken wavelength coverage providing many hydrogen lines with high S/N and sufficient spectral resolution, there are many existing archival Herbig AeBe spectra from instruments such as x-shooter that are well suited for this type of study. Some potential challenges with spectra from ground based facilities include gaps in wavelength coverage due to telluric absorption in the NIR, larger extinction correction uncertainties at optical wavelengths, and the importance of other line broadening mechanisms for certain Balmer lines making the kinematic analysis less straightforward. Despite these obstacles, the large sample of archival spectra provides an opportunity to expand this study with a statistically significant sample.
\section{Conclusions}
\label{sec:conclusions}
We have presented JWST NIRSpec spectra of five intermediate mass sources located in the massive star forming region NGC 3603. The spectra of our sources exhibit many recombination lines (mostly from H I) and lack any detectable absorption lines from their underlying photospheres. We fit their optical photometry with an MCMC exploration to estimate their stellar properties. We have classified sources 185, 238, and 823 as Herbig Ae stars based on the best-fitting $T_{eff}$. Sources 251 and 469 have spectra consistent with spectral types G and F, respectively. Sources 185, 238, and 251 exhibit CO bandhead emission. We performed a kinematic analysis on the multiple series of hydrogen emission lines to try to constrain where the line emission originates. Based on the FWHM and optical depth of the spectrally resolved emission lines and considering the typical velocities expected for a number line emission scenarios, we favour an origin in the MA flow for sources 185, 238, and 823. We cannot rule out the possibility of emission from the Keplerian disc for sources 251 and 469. Surprisingly, none of our sources are consistent with emission from a magneto-centrifugal wind, despite this being a commonly invoked mechanism to explain hydrogen line emission for Herbig AeBe stars. This result provides new observational support to the existence of a magnetically driven accretion mechanism around Herbig Ae stars, which can dominate the line emission from these sources. With a small sample size of five, it is unwise to draw generalised conclusions from these sources alone. The kinematic analysis that we have presented and described here can be extended to the large sample of existing archival Herbig AeBe spectra, provided enough hydrogen lines are present with sufficient S/N and spectral resolution. Increasing the sample size would provide insight into the extent to which MA dominates the line emission for Herbig AeBe stars.

\begin{acknowledgements}
      We wish to thank the referee for their careful review of this work. Their comments and suggestions significantly improved the quality of our analysis.
{\em Software}: 
\texttt{SciPy} \citep{virtanen2020scipy};
\texttt{NumPy} \citep{harris2020array};
\texttt{Matplotlib} \citep{hunter2007matplotlib}; 
\texttt{MSAfit} \citep{de2024ionised}

\end{acknowledgements}

\bibliographystyle{aa} 
\bibliography{references.bib} 

\begin{thebibliography}{83}
\expandafter\ifx\csname natexlab\endcsname\relax\def\natexlab#1{#1}\fi

\bibitem[{Abbott(1982)}]{abbott1982theory}
Abbott, D.~C. 1982, Astrophysical Journal, Part 1, vol. 259, Aug. 1, 1982, p. 282-301., 259, 282

\bibitem[{Alcal{\'a} {et~al.}(2021)Alcal{\'a}, Gangi, Biazzo, Antoniucci, Frasca, Giannini, Munari, Nisini, Harutyunyan, Manara, {et~al.}}]{alcala2021giarps}
Alcal{\'a}, J., Gangi, M., Biazzo, K., {et~al.} 2021, Astronomy \& Astrophysics, 652, A72

\bibitem[{Alcal{\'a} {et~al.}(2017)Alcal{\'a}, Manara, Natta, Frasca, Testi, Nisini, Stelzer, Williams, Antoniucci, Biazzo, {et~al.}}]{alcala2017x}
Alcal{\'a}, J., Manara, C., Natta, A., {et~al.} 2017, Astronomy \& Astrophysics, 600, A20

\bibitem[{Alcal{\'a} {et~al.}(2014)Alcal{\'a}, Natta, Manara, Spezzi, Stelzer, Frasca, Biazzo, Covino, Randich, Rigliaco, {et~al.}}]{alcala2014x}
Alcal{\'a}, J., Natta, A., Manara, C., {et~al.} 2014, Astronomy \& Astrophysics, 561, A2

\bibitem[{Antoniucci {et~al.}(2017)Antoniucci, Nisini, Biazzo, Giannini, Lorenzetti, Sanna, Harutyunyan, Origlia, \& Oliva}]{antoniucci2017high}
Antoniucci, S., Nisini, B., Biazzo, K., {et~al.} 2017, Astronomy \& Astrophysics, 606, A48

\bibitem[{Baker \& Menzel(1938)}]{baker1938physical}
Baker, J.~G. \& Menzel, D.~H. 1938, Astrophysical Journal, vol. 88, p. 52, 88, 52

\bibitem[{Beccari {et~al.}(2010)Beccari, Spezzi, De~Marchi, Paresce, Young, Andersen, Panagia, Balick, Bond, Calzetti, {et~al.}}]{beccari2010progressive}
Beccari, G., Spezzi, L., De~Marchi, G., {et~al.} 2010, The Astrophysical Journal, 720, 1108

\bibitem[{Bell {et~al.}(2013)Bell, Naylor, Mayne, Jeffries, \& Littlefair}]{bell2013pre}
Bell, C.~P., Naylor, T., Mayne, N., Jeffries, R., \& Littlefair, S. 2013, Monthly Notices of the Royal Astronomical Society, 434, 806

\bibitem[{Blandford \& Payne(1982)}]{blandford1982hydromagnetic}
Blandford, R.~D. \& Payne, D. 1982, Monthly Notices of the Royal Astronomical Society, 199, 883

\bibitem[{B{\"o}hm \& Catala(1995)}]{bohm1995rotation}
B{\"o}hm, T. \& Catala, C. 1995, Astronomy and Astrophysics, v. 301, p. 155, 301, 155

\bibitem[{Bouvier {et~al.}(2007)Bouvier, Alencar, Boutelier, Dougados, Balog, Grankin, Hodgkin, Ibrahimov, Kun, Magakian, {et~al.}}]{bouvier2007magnetospheric}
Bouvier, J., Alencar, S., Boutelier, T., {et~al.} 2007, Astronomy \& Astrophysics, 463, 1017

\bibitem[{Bunn {et~al.}(1995)Bunn, Hoare, \& Drew}]{bunn1995observations}
Bunn, J., Hoare, M., \& Drew, J. 1995, Monthly Notices of the Royal Astronomical Society, 272, 346

\bibitem[{Cauley \& Johns-Krull(2014)}]{cauley2014diagnosing}
Cauley, P.~W. \& Johns-Krull, C.~M. 2014, The Astrophysical Journal, 797, 112

\bibitem[{Cauley \& Johns-Krull(2015)}]{cauley2015optical}
Cauley, P.~W. \& Johns-Krull, C.~M. 2015, The Astrophysical Journal, 810, 5

\bibitem[{Cieza {et~al.}(2005)Cieza, Kessler-Silacci, Jaffe, Harvey, \& Evans~II}]{cieza2005evidence}
Cieza, L.~A., Kessler-Silacci, J.~E., Jaffe, D.~T., Harvey, P.~M., \& Evans~II, N.~J. 2005, The Astrophysical Journal, 635, 422

\bibitem[{Czekala {et~al.}(2015)Czekala, Andrews, Mandel, Hogg, \& Green}]{czekala2015constructing}
Czekala, I., Andrews, S.~M., Mandel, K.~S., Hogg, D.~W., \& Green, G.~M. 2015, The Astrophysical Journal, 812, 128

\bibitem[{de~Graaff {et~al.}(2024)de~Graaff, Rix, Carniani, Suess, Charlot, Curtis-Lake, Arribas, Baker, Boyett, Bunker, {et~al.}}]{de2024ionised}
de~Graaff, A., Rix, H.-W., Carniani, S., {et~al.} 2024, Astronomy \& Astrophysics, 684, A87

\bibitem[{Dimitrijevic(2007)}]{dimitrijevic2007stark}
Dimitrijevic, M. 2007, in Solar and Stellar Physics Through Eclipses, Vol. 370, 270

\bibitem[{Dorner {et~al.}(2016)Dorner, Giardino, Ferruit, de~Oliveira, Birkmann, B{\"o}ker, De~Marchi, Gnata, K{\"o}hler, Sirianni, {et~al.}}]{NIPS}
Dorner, B., Giardino, G., Ferruit, P., {et~al.} 2016, Astronomy \& Astrophysics, 592, A113

\bibitem[{Drew {et~al.}(2019)Drew, Mongui{\'o}, \& Wright}]{drew2019star}
Drew, J., Mongui{\'o}, M., \& Wright, N. 2019, Monthly Notices of the Royal Astronomical Society, 486, 1034

\bibitem[{Drissen {et~al.}(1995)Drissen, Moffat, Walborn, \& Shara}]{drissen1995dense}
Drissen, L., Moffat, A.~F., Walborn, N.~R., \& Shara, M.~M. 1995, Astronomical Journal v. 110, p. 2235, 110, 2235

\bibitem[{{Dupree} {et~al.}(2005){Dupree}, {Brickhouse}, {Smith}, \& {Strader}}]{2005ApJ...625L.131D}
{Dupree}, A.~K., {Brickhouse}, N.~S., {Smith}, G.~H., \& {Strader}, J. 2005, \apjl, 625, L131

\bibitem[{Edwards {et~al.}(2006)Edwards, Fischer, Hillenbrand, \& Kwan}]{edwards2006probing}
Edwards, S., Fischer, W., Hillenbrand, L., \& Kwan, J. 2006, The Astrophysical Journal, 646, 319

\bibitem[{Ercolano \& Owen(2010)}]{ercolano2010theoretical}
Ercolano, B. \& Owen, J.~E. 2010, Monthly Notices of the Royal Astronomical Society, 406, 1553

\bibitem[{Fairlamb {et~al.}(2015)Fairlamb, Oudmaijer, Mendigutia, Ilee, \& van~den Ancker}]{fairlamb2015spectroscopic}
Fairlamb, J.~R., Oudmaijer, R.~D., Mendigutia, I., Ilee, J.~D., \& van~den Ancker, M.~E. 2015, Monthly Notices of the Royal Astronomical Society, 453, 976

\bibitem[{Feldman \& Doschek(1977)}]{feldman1977emission}
Feldman, U. \& Doschek, G. 1977, Astrophysical Journal, Vol. 212, pp. 913-922 (1977)., 212, 913

\bibitem[{Ferruit {et~al.}(2022)Ferruit, Jakobsen, Giardino, Rawle, de~Oliveira, Arribas, Beck, Birkmann, B{\"o}ker, Bunker, {et~al.}}]{ferruit2022near}
Ferruit, P., Jakobsen, P., Giardino, G., {et~al.} 2022, Astronomy \& Astrophysics, 661, A81

\bibitem[{Frank {et~al.}(2014)Frank, Ray, Cabrit, Hartigan, Arce, Bacciotti, Bally, Benisty, Eisl{\"o}ffel, G{\"u}del, {et~al.}}]{frank2014jets}
Frank, A., Ray, T., Cabrit, S., {et~al.} 2014, Protostars and planets VI, 451

\bibitem[{Gordon {et~al.}(2022)Gordon, Bohlin, Sloan, Rieke, Volk, Boyer, Muzerolle, Schlawin, Deustua, Hines, {et~al.}}]{gordon2022james}
Gordon, K.~D., Bohlin, R., Sloan, G., {et~al.} 2022, The Astronomical Journal, 163, 267

\bibitem[{Gordon {et~al.}(2023)Gordon, Clayton, Decleir, Fitzpatrick, Massa, Misselt, \& Tollerud}]{gordon2023one}
Gordon, K.~D., Clayton, G.~C., Decleir, M., {et~al.} 2023, The Astrophysical Journal, 950, 86

\bibitem[{Gullbring {et~al.}(1998)Gullbring, Hartmann, Briceno, \& Calvet}]{gullbring1998disk}
Gullbring, E., Hartmann, L., Briceno, C., \& Calvet, N. 1998, The Astrophysical Journal, 492, 323

\bibitem[{Hamaguchi {et~al.}(2005)Hamaguchi, Yamauchi, \& Koyama}]{hamaguchi2005x}
Hamaguchi, K., Yamauchi, S., \& Koyama, K. 2005, The Astrophysical Journal, 618, 360

\bibitem[{Harris {et~al.}(2020)Harris, Millman, Van Der~Walt, Gommers, Virtanen, Cournapeau, Wieser, Taylor, Berg, Smith, {et~al.}}]{harris2020array}
Harris, C.~R., Millman, K.~J., Van Der~Walt, S.~J., {et~al.} 2020, Nature, 585, 357

\bibitem[{Harsono {et~al.}(2023)Harsono, Bjerkeli, Ramsey, Pontoppidan, Kristensen, J{\o}rgensen, Calcutt, Li, \& Plunkett}]{harsono2023jwst}
Harsono, D., Bjerkeli, P., Ramsey, J., {et~al.} 2023, The Astrophysical Journal Letters, 951, L32

\bibitem[{Hartmann {et~al.}(2016)Hartmann, Herczeg, \& Calvet}]{hartmann2016accretion}
Hartmann, L., Herczeg, G., \& Calvet, N. 2016, Annual Review of Astronomy and Astrophysics, 54, 135

\bibitem[{Herczeg \& Hillenbrand(2008)}]{herczeg2008uv}
Herczeg, G.~J. \& Hillenbrand, L.~A. 2008, The Astrophysical Journal, 681, 594

\bibitem[{Hunter \& Dale(2007)}]{hunter2007matplotlib}
Hunter, J. \& Dale, D. 2007, Matplotlib 0.90. 0 user’s guide, 487

\bibitem[{Husser {et~al.}(2013)Husser, Wende-von Berg, Dreizler, Homeier, Reiners, Barman, \& Hauschildt}]{husser2013new}
Husser, T.-O., Wende-von Berg, S., Dreizler, S., {et~al.} 2013, Astronomy \& Astrophysics, 553, A6

\bibitem[{Ilee {et~al.}(2013)Ilee, Wheelwright, Oudmaijer, de~Wit, Maud, Hoare, Lumsden, Moore, Urquhart, \& Mottram}]{ilee2013co}
Ilee, J., Wheelwright, H., Oudmaijer, R., {et~al.} 2013, Monthly Notices of the Royal Astronomical Society, 429, 2960

\bibitem[{Johns-Krull(2007)}]{johns2007magnetic}
Johns-Krull, C.~M. 2007, The Astrophysical Journal, 664, 975

\bibitem[{Kageyama \& Sato(1997)}]{kageyama1997generation}
Kageyama, A. \& Sato, T. 1997, Physical review E, 55, 4617

\bibitem[{Koenigl(1991)}]{koenigl1991disk}
Koenigl, A. 1991, Astrophysical Journal, Part 2-Letters (ISSN 0004-637X), vol. 370, March 20, 1991, p. L39-L43. Research supported by Rockwell International Corp. and Illinois Space Institute., 370, L39

\bibitem[{Kraus {et~al.}(2008)Kraus, Hofmann, Benisty, Berger, Chesneau, Isella, Malbet, Meilland, Nardetto, Natta, {et~al.}}]{kraus2008origin}
Kraus, S., Hofmann, K.-H., Benisty, M., {et~al.} 2008, Astronomy \& Astrophysics, 489, 1157

\bibitem[{Kreplin {et~al.}(2018)Kreplin, Tambovtseva, Grinin, Kraus, Weigelt, \& Wang}]{kreplin2018brgamma}
Kreplin, A., Tambovtseva, L., Grinin, V., {et~al.} 2018, Monthly Notices of the Royal Astronomical Society, 476, 4520

\bibitem[{Kudritzki \& Puls(2000)}]{kudritzki2000winds}
Kudritzki, R.-P. \& Puls, J. 2000, Annual Review of Astronomy and Astrophysics, 38, 613

\bibitem[{Kuncarayakti {et~al.}(2016)Kuncarayakti, Galbany, Anderson, Kr{\"u}hler, \& Hamuy}]{kuncarayakti2016unresolved}
Kuncarayakti, H., Galbany, L., Anderson, J., Kr{\"u}hler, T., \& Hamuy, M. 2016, Astronomy \& Astrophysics, 593, A78

\bibitem[{Kurosawa {et~al.}(2006)Kurosawa, Harries, \& Symington}]{kurosawa2006formation}
Kurosawa, R., Harries, T.~J., \& Symington, N.~H. 2006, Monthly Notices of the Royal Astronomical Society, 370, 580

\bibitem[{Lima {et~al.}(2010)Lima, Alencar, Calvet, Hartmann, \& Muzerolle}]{lima2010modeling}
Lima, G., Alencar, S., Calvet, N., Hartmann, L., \& Muzerolle, J. 2010, Astronomy \& Astrophysics, 522, A104

\bibitem[{Lumsden {et~al.}(2012)Lumsden, Wheelwright, Hoare, Oudmaijer, \& Drew}]{lumsden2012tracers}
Lumsden, S., Wheelwright, H., Hoare, M., Oudmaijer, R., \& Drew, J. 2012, Monthly Notices of the Royal Astronomical Society, 424, 1088

\bibitem[{Lynden-Bell \& Pringle(1974)}]{lynden1974evolution}
Lynden-Bell, D. \& Pringle, J.~E. 1974, Monthly Notices of the Royal Astronomical Society, 168, 603

\bibitem[{McClure {et~al.}(2013)McClure, Calvet, Espaillat, Hartmann, Hern{\'a}ndez, Ingleby, Luhman, D'Alessio, \& Sargent}]{mcclure2013characterizing}
McClure, M., Calvet, N., Espaillat, C., {et~al.} 2013, The Astrophysical Journal, 769, 73

\bibitem[{Mendigut{\'\i}a(2020)}]{mendigutia2020mass}
Mendigut{\'\i}a, I. 2020, Galaxies, 8, 39

\bibitem[{Mendigut{\'\i}a {et~al.}(2015)Mendigut{\'\i}a, de~Wit, Oudmaijer, Fairlamb, Carciofi, Ilee, \& Vieira}]{mendigutia2015high}
Mendigut{\'\i}a, I., de~Wit, W., Oudmaijer, R., {et~al.} 2015, Monthly Notices of the Royal Astronomical Society, 453, 2126

\bibitem[{Mendigut{\'\i}a {et~al.}(2017)Mendigut{\'\i}a, Oudmaijer, Mourard, \& Muzerolle}]{mendigutia2017compact}
Mendigut{\'\i}a, I., Oudmaijer, R., Mourard, D., \& Muzerolle, J. 2017, Monthly Notices of the Royal Astronomical Society, 464, 1984

\bibitem[{Moffat {et~al.}(1994)Moffat, Drissen, \& Shara}]{moffat1994ngc}
Moffat, A.~F., Drissen, L., \& Shara, M.~M. 1994, Astrophysical Journal, Part 1 (ISSN 0004-367X), vol. 436, no. 1, p. 183-193, 436, 183

\bibitem[{Muzerolle {et~al.}(1998{\natexlab{a}})Muzerolle, Calvet, \& Hartmann}]{muzerolle1998magnetospheric}
Muzerolle, J., Calvet, N., \& Hartmann, L. 1998{\natexlab{a}}, The Astrophysical Journal, 492, 743

\bibitem[{Muzerolle {et~al.}(2003)Muzerolle, Calvet, Hartmann, \& D’Alessio}]{muzerolle2003unveiling}
Muzerolle, J., Calvet, N., Hartmann, L., \& D’Alessio, P. 2003, The Astrophysical Journal, 597, L149

\bibitem[{Muzerolle {et~al.}(2004)Muzerolle, D’Alessio, Calvet, \& Hartmann}]{muzerolle2004magnetospheres}
Muzerolle, J., D’Alessio, P., Calvet, N., \& Hartmann, L. 2004, The Astrophysical Journal, 617, 406

\bibitem[{Muzerolle {et~al.}(1998{\natexlab{b}})Muzerolle, Hartmann, \& Calvet}]{muzerolle1998brgamma}
Muzerolle, J., Hartmann, L., \& Calvet, N. 1998{\natexlab{b}}, The Astronomical Journal, 116, 2965

\bibitem[{Nisini {et~al.}(2004)Nisini, Antoniucci, \& Giannini}]{nisini2004observations}
Nisini, B., Antoniucci, S., \& Giannini, T. 2004, Astronomy \& Astrophysics, 421, 187

\bibitem[{Owen {et~al.}(2012)Owen, Clarke, \& Ercolano}]{owen2012theory}
Owen, J.~E., Clarke, C.~J., \& Ercolano, B. 2012, Monthly Notices of the Royal Astronomical Society, 422, 1880

\bibitem[{Popovi{\'c} {et~al.}(2001)Popovi{\'c}, Simi{\'c}, Milovanovi{\'c}, \& Dimitrijevi{\'c}}]{popovic2001stark}
Popovi{\'c}, L., Simi{\'c}, S., Milovanovi{\'c}, N., \& Dimitrijevi{\'c}, M. 2001, The Astrophysical Journal Supplement Series, 135, 109

\bibitem[{Reipurth \& Bally(2001)}]{reipurth2001herbig}
Reipurth, B. \& Bally, J. 2001, Annual Review of Astronomy and Astrophysics, 39, 403

\bibitem[{Reipurth {et~al.}(1998)Reipurth, Bally, Fesen, \& Devine}]{reipurth1998protostellar}
Reipurth, B., Bally, J., Fesen, R.~A., \& Devine, D. 1998, Nature, 396, 343

\bibitem[{Reiter {et~al.}(2024)Reiter, Haworth, Manara, Ramsay, Klaassen, Itrich, \& McLeod}]{reiter2024illuminating}
Reiter, M., Haworth, T.~J., Manara, C.~F., {et~al.} 2024, Monthly Notices of the Royal Astronomical Society, 527, 3220

\bibitem[{Reiter {et~al.}(2022)Reiter, Morse, Smith, Haworth, Kuhn, \& Klaassen}]{reiter2022deep}
Reiter, M., Morse, J.~A., Smith, N., {et~al.} 2022, Monthly Notices of the Royal Astronomical Society, 517, 5382

\bibitem[{Reiter \& Smith(2013)}]{reiter2013hst}
Reiter, M. \& Smith, N. 2013, Monthly Notices of the Royal Astronomical Society, 433, 2226

\bibitem[{Reiter {et~al.}(2016)Reiter, Smith, \& Bally}]{reiter2016fe}
Reiter, M., Smith, N., \& Bally, J. 2016, Monthly Notices of the Royal Astronomical Society, 463, 4344

\bibitem[{Ricci {et~al.}(2008)Ricci, Robberto, \& Soderblom}]{ricci2008hubble}
Ricci, L., Robberto, M., \& Soderblom, D.~R. 2008, The Astronomical Journal, 136, 2136

\bibitem[{Rogers {et~al.}(2024{\natexlab{a}})Rogers, Brandl, \& De~Marchi}]{rogers2024spectral}
Rogers, C., Brandl, B., \& De~Marchi, G. 2024{\natexlab{a}}, Astronomy \& Astrophysics, 688, A111

\bibitem[{Rogers {et~al.}(2024{\natexlab{b}})Rogers, de~Marchi, \& Brandl}]{rogers2024determining}
Rogers, C., de~Marchi, G., \& Brandl, B. 2024{\natexlab{b}}, Astronomy \& Astrophysics, 684, L8

\bibitem[{Rogers {et~al.}(2024{\natexlab{c}})Rogers, de~Marchi, \& Brandl}]{RogersCTTS}
Rogers, C., de~Marchi, G., \& Brandl, B. 2024{\natexlab{c}}, Externally irradiated young stars in NGC 3603. A JWST NIRSpec catalogue of pre-main-sequence stars in a massive star formation region

\bibitem[{Storey \& Hummer(1995)}]{storey1995recombination}
Storey, P. \& Hummer, D. 1995, Monthly Notices of the Royal Astronomical Society, 272, 41

\bibitem[{Tabone {et~al.}(2022)Tabone, Rosotti, Lodato, Armitage, Cridland, \& van Dishoeck}]{tabone2022mhd}
Tabone, B., Rosotti, G.~P., Lodato, G., {et~al.} 2022, Monthly Notices of the Royal Astronomical Society: Letters, 512, L74

\bibitem[{Tambovtseva {et~al.}(2014)Tambovtseva, Grinin, \& Weigelt}]{tambovtseva2014hydrogen}
Tambovtseva, L., Grinin, V., \& Weigelt, G. 2014, Astronomy \& Astrophysics, 562, A104

\bibitem[{Tambovtseva {et~al.}(2016)Tambovtseva, Grinin, \& Weigelt}]{tambovtseva2016brackett}
Tambovtseva, L., Grinin, V., \& Weigelt, G. 2016, Astronomy \& Astrophysics, 590, A97

\bibitem[{Vioque {et~al.}(2022)Vioque, Oudmaijer, Wichittanakom, Mendigut{\'\i}a, Baines, Pani{\'c}, Iglesias, Miley, \& P{\'e}rez-Mart{\'\i}nez}]{vioque2022identification}
Vioque, M., Oudmaijer, R.~D., Wichittanakom, C., {et~al.} 2022, The Astrophysical Journal, 930, 39

\bibitem[{Virtanen {et~al.}(2020)Virtanen, Gommers, Oliphant, Haberland, Reddy, Cournapeau, Burovski, Peterson, Weckesser, Bright, {et~al.}}]{virtanen2020scipy}
Virtanen, P., Gommers, R., Oliphant, T.~E., {et~al.} 2020, Nature methods, 17, 261

\bibitem[{Wichittanakom {et~al.}(2020)Wichittanakom, Oudmaijer, Fairlamb, Mendigut{\'\i}a, Vioque, \& Ababakr}]{wichittanakom2020accretion}
Wichittanakom, C., Oudmaijer, R., Fairlamb, J., {et~al.} 2020, Monthly Notices of the Royal Astronomical Society, 493, 234

\bibitem[{Williams \& Cieza(2011)}]{williams2011protoplanetary}
Williams, J.~P. \& Cieza, L.~A. 2011, Annual Review of Astronomy and Astrophysics, 49, 67

\bibitem[{Wilson {et~al.}(2022)Wilson, Matt, Harries, \& Herczeg}]{wilson2022hydrogen}
Wilson, T.~J., Matt, S., Harries, T., \& Herczeg, G. 2022, Monthly Notices of the Royal Astronomical Society, 514, 2162

\bibitem[{Winter \& Haworth(2022)}]{winter2022external}
Winter, A.~J. \& Haworth, T.~J. 2022, The European Physical Journal Plus, 137, 1132

\bibitem[{Zinnecker \& Preibisch(1994)}]{zinnecker1994x}
Zinnecker, H. \& Preibisch, T. 1994, Astronomy and Astrophysics (ISSN 0004-6361), vol. 292, no. 1, p. 152-164, 292, 152

\end{thebibliography}
\begin{appendix}
\section{FWHM diagrams} \label{subsec:FWHM_diagrams}
\begin{figure*}[hbt!]
    \centering
    \onecolumn 
    \includegraphics[width=1\linewidth]{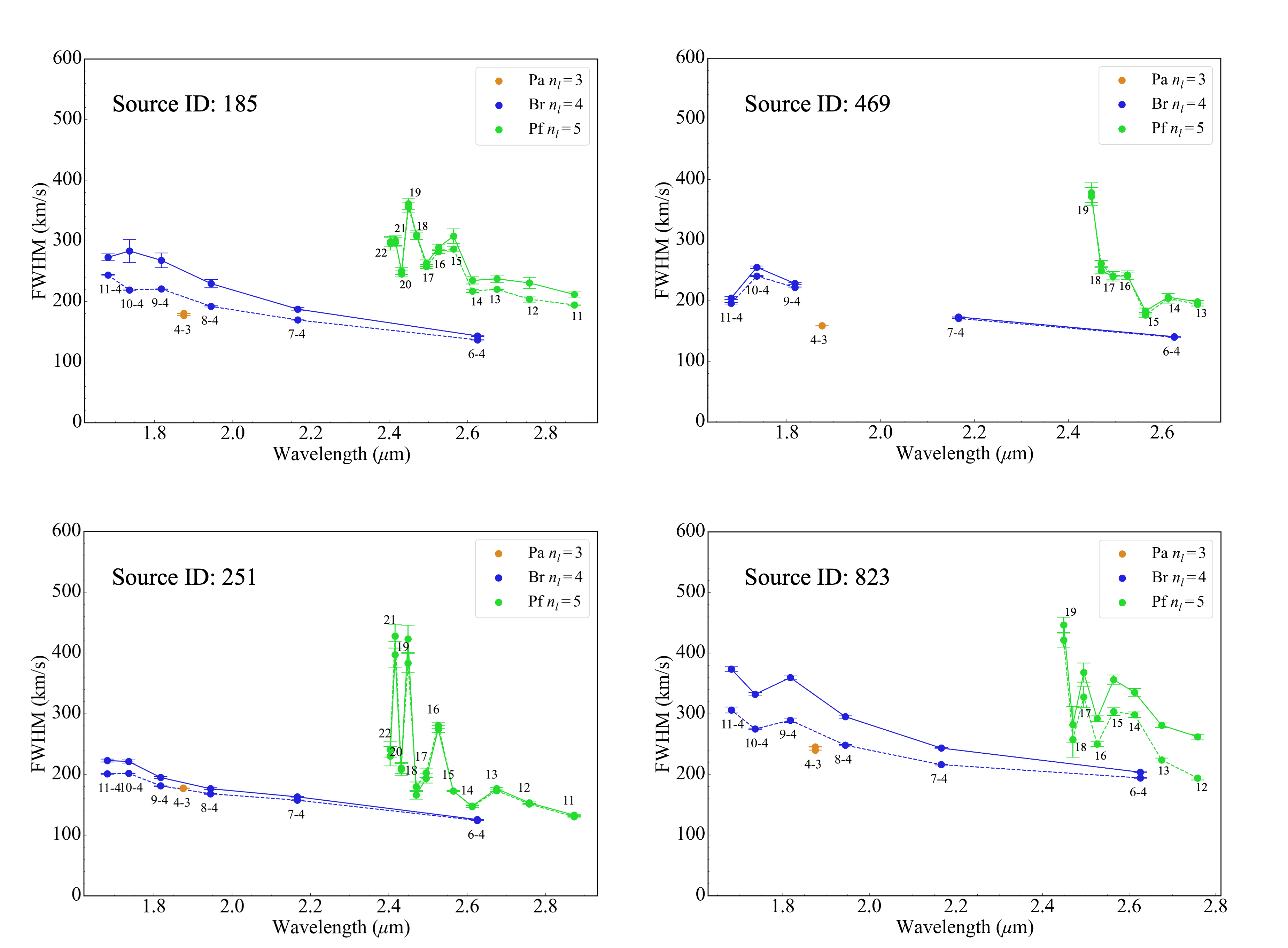}
    \caption{Full width at half maximum of each hydrogen emission line. Top left - Source 185. Bottom left - Source 251. Top right - Source 469. Bottom right - Source 823.}
    \label{fig:FWHM_diagrams}
\end{figure*}

\newpage

\section{Optical depth diagrams} \label{subsec:optical_depth_appendix}
\begin{figure*}[hbt!]
    \centering
    \onecolumn
    \includegraphics[width=0.75\linewidth]{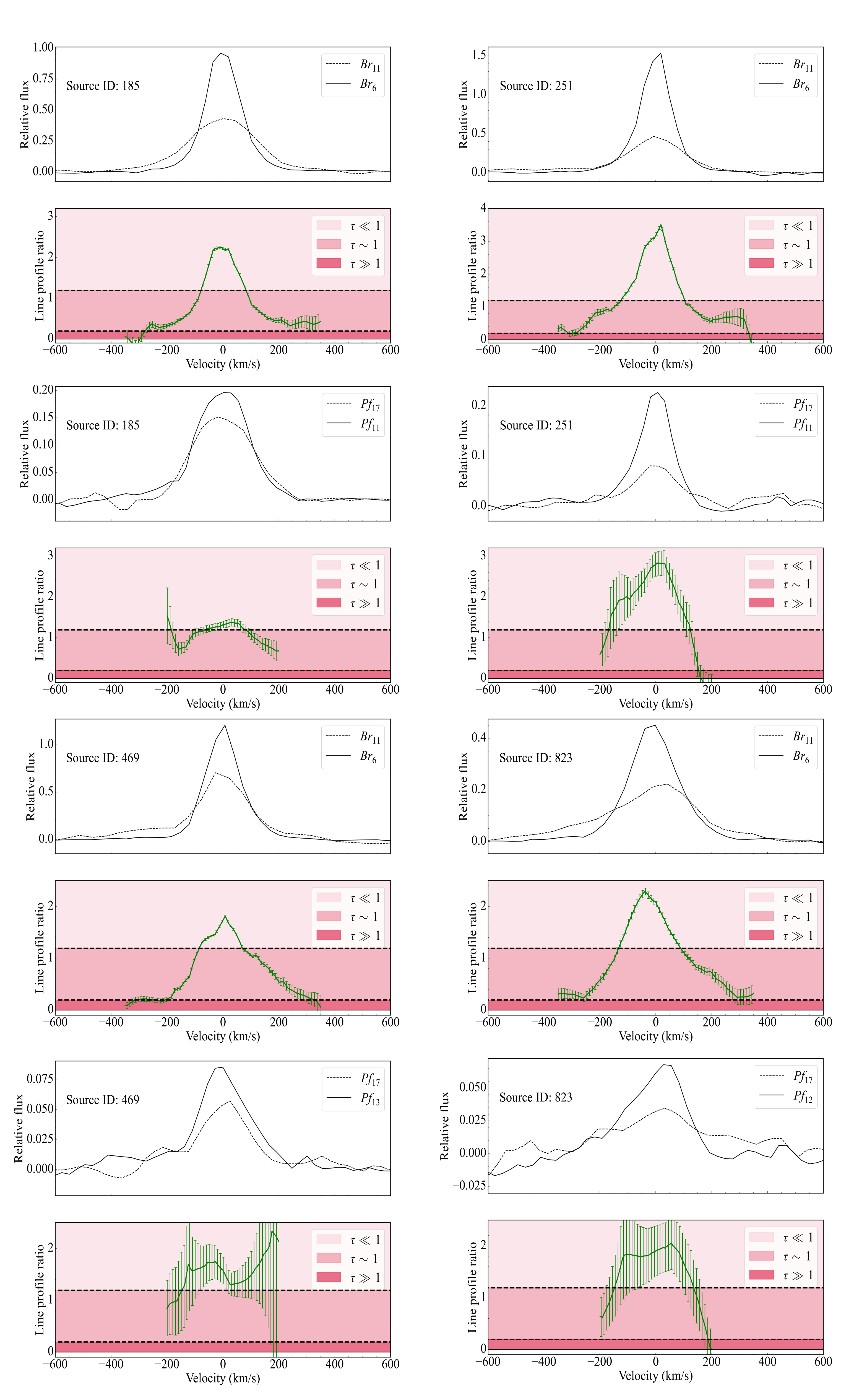}
    \caption{Optical depth diagrams of sources 185, 251, 469, and 823. In the case of source 469, $Pf_{13}$ was used, as neither $Pf_{12}$ or $Pf_{11}$ were measured. In the case of source 823, $Pf_{12}$ was used, as $Pf_{11}$ was not measured.}
    \label{fig:optical_depth_grid}
\end{figure*}

\clearpage
\section{MUSE spectra best fits} \label{MUSE_spectra}

\begin{figure*}[hbt!]
    \centering
    \onecolumn
    \includegraphics[width=0.75\linewidth]{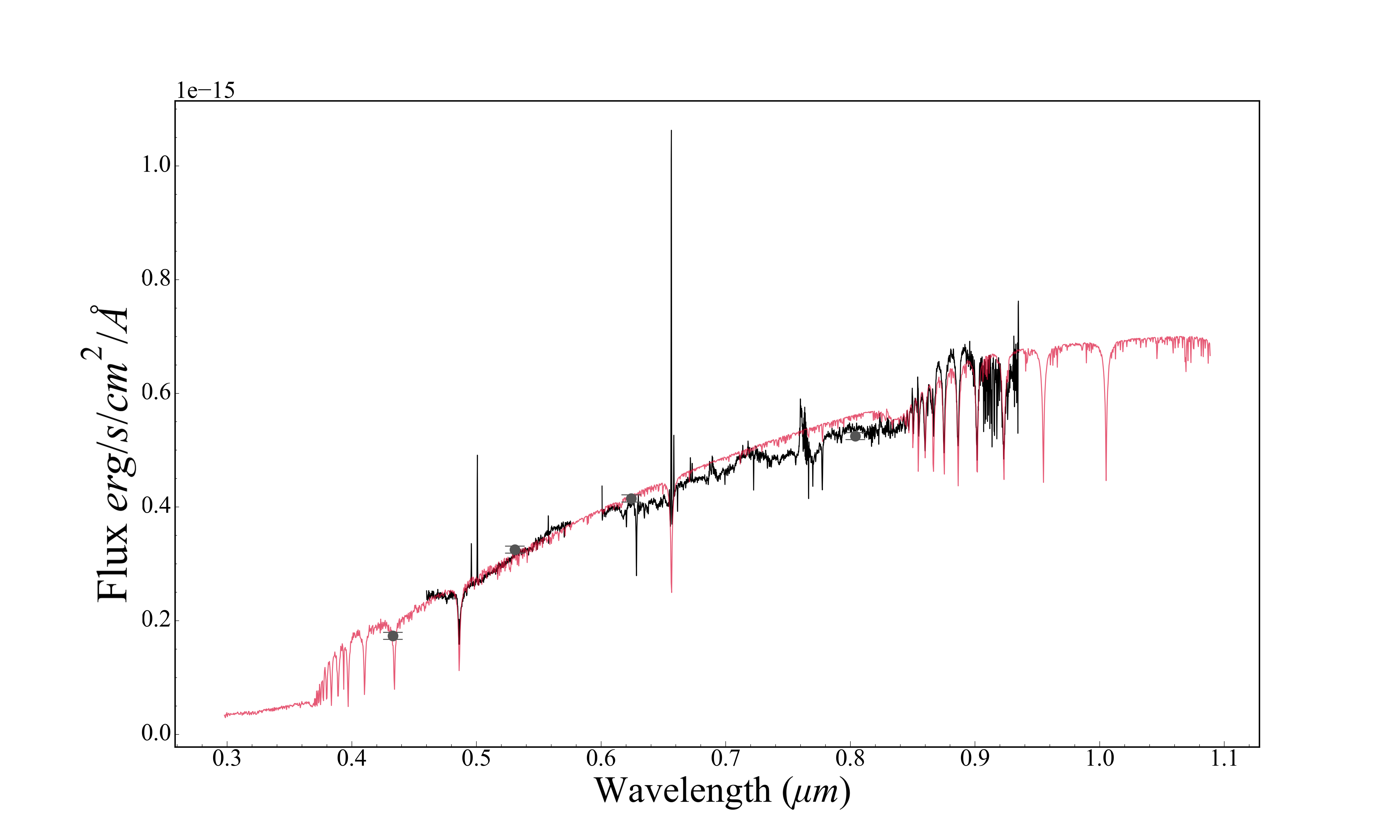}
    \caption{Best-fitting optical model spectrum for source $152$ based on its MUSE spectrum. In black is the MUSE source spectrum. In red is the best-fitting model spectrum. In grey is the HST photometry.}
    \label{fig:152}
\end{figure*}

\begin{figure*}[hbt!]
    \centering
    \onecolumn
    \includegraphics[width=0.75\linewidth]{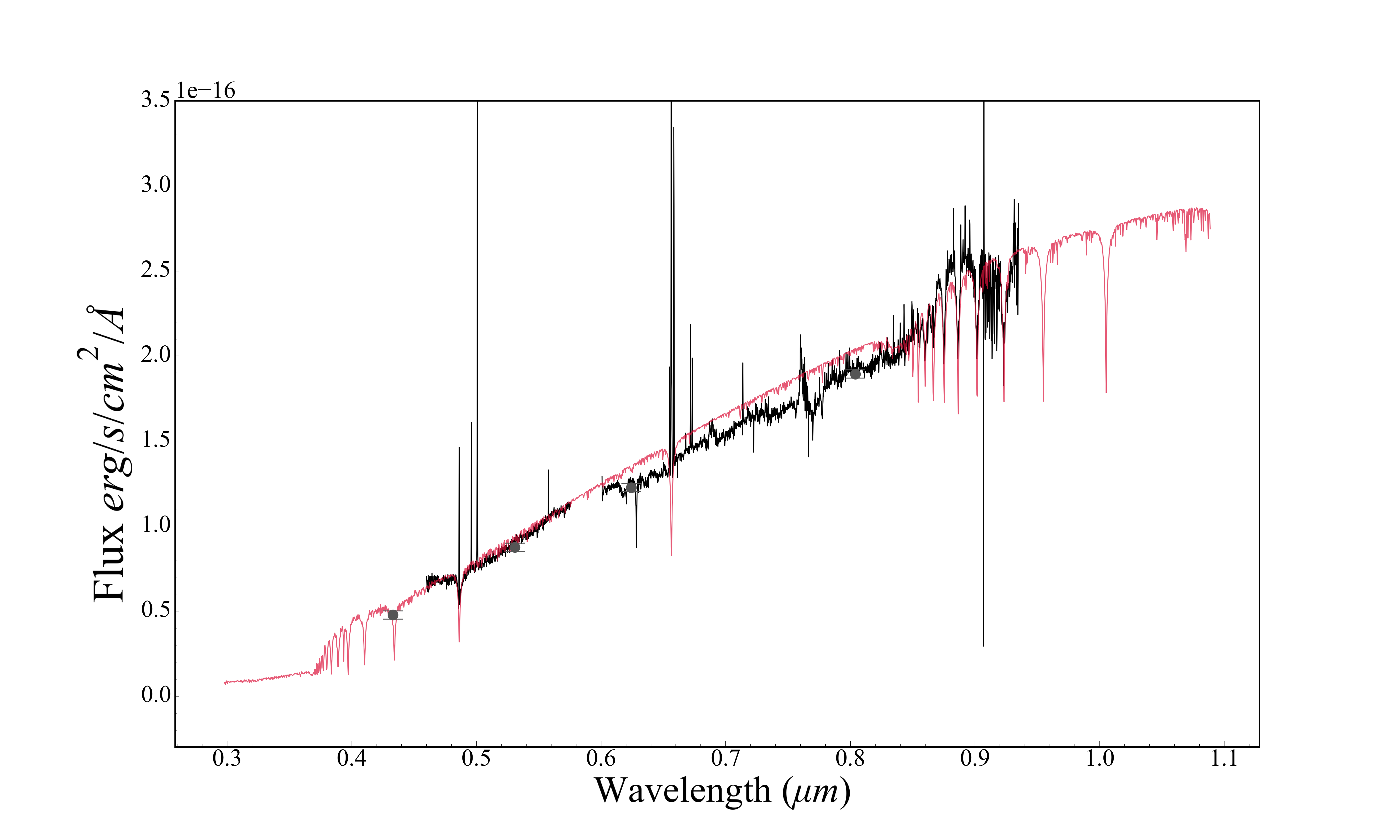}
    \caption{As before but for source $354$.}
    \label{fig:352}
\end{figure*}

\newpage

\section{Emission line properties} \label{emission_line_tables}

\setlength{\tabcolsep}{4pt} 
\begin{table*}[h]
    \centering
    \caption{Emission line properties for sources 185, 238, and 251.}
    \label{tab:combined}
    \begin{tabular}{l c c c c c c}
        \toprule
        & \multicolumn{2}{c}{\textbf{Source 185}} & \multicolumn{2}{c}{\textbf{Source 238}} & \multicolumn{2}{c}{\textbf{Source 251}} \\
        \cmidrule(lr){2-3} \cmidrule(lr){4-5} \cmidrule(lr){6-7} 
        Parameter & EW (\AA) & Luminosity ($L_{\odot}$) & EW (\AA) & Luminosity ($L_{\odot}$) & EW (\AA) & Luminosity ($L_{\odot}$) \\
        \midrule
        Pa$_\alpha$  & $-79.198 \pm \; 7.316$ & $-0.146 \pm \; 0.0141$ & $-74.163 \pm \; 5.517$ & $-0.07 \pm \; 0.0056$ & $-130.731 \pm \; 9.089 $ & $-0.295 \pm \; 0.022$ \\
        Br$_{6-4}$   & $-13.845 \pm \; 1.271$ & $-0.020 \pm \; 0.0019$ & $-24.657 \pm \; 1.852$ & $-0.013 \pm \; 0.001$ & $-18.944 \pm \; 1.305$ & $-0.05 \pm \; 0.004$ \\
        Br$_{7-4}$   & $-10.460 \pm \; 0.974$ & $-0.018 \pm \; 0.0017$ & $-16.408 \pm \; 1.323$ & $-0.012 \pm \; 0.0011$ & $-15.104 \pm \; 1.049$ & $-0.034 \pm \; 0.003$ \\
        Br$_{8-4}$  & $-8.592 \pm \; 0.864$ & $-0.017 \pm \; 0.0017$  & $-15.143 \pm \; 1.215$ & $-0.014 \pm \; 0.0012$ & $-12.006 \pm \; 0.827$ & $-0.027 \pm \; 0.002$ \\
        Br$_{9-4}$   & $-8.921 \pm \; 1.013$ & $-0.017 \pm \; 0.0019$  & $-14.425 \pm \; 1.201$ & $-0.014 \pm \; 0.0012$ & $-9.222 \pm \; 0.643$ & $-0.021 \pm \; 0.002$ \\
        Br$_{10-4}$  & $-8.313 \pm \; 1.035$ & $-0.016 \pm \; 0.002$  & $-15.303 \pm \; 1.282$ & $-0.015 \pm \; 0.0013$ & $-8.372 \pm \; 0.619$ & $-0.019 \pm \; 0.002$ \\
        Br$_{11-4}$  & $-7.338 \pm \; 0.781$ & $-0.014 \pm \; 0.0015$  & $-13.331 \pm \; 1.205$ & $-0.014 \pm \; 0.0013$ & $-6.514 \pm \; 0.463$ & $-0.015 \pm \; 0.0011$ \\
        Pf$_{11-5}$  & $-4.485 \pm \; 0.454$ & $-0.006 \pm \; 0.0006$  & $-14.787 \pm \; 1.137$ & $-0.007 \pm \; 0.0005$ & $-3.225 \pm \; 0.224$ & $-0.009 \pm \; 0.0006$ \\
        Pf$_{12-5}$  & $-4.464 \pm \; 0.498$ & $-0.006 \pm \; 0.0007$  & $-15.607 \pm \; 1.215$ & $-0.007 \pm \; 0.0006$ & $-2.665 \pm \; 0.187$ & $-0.007 \pm \; 0.0005$ \\
        Pf$_{13-5}$  & $-4.402 \pm \; 0.460$ & $-0.006 \pm \; 0.0007$  & $-10.357 \pm \; 0.816$ & $-0.005 \pm \; 0.0004$ & $-2.563 \pm \; 0.173$ & $-0.006 \pm \; 0.0005$ \\
        Pf$_{14-5}$  & $-4.250 \pm \; 0.432$ & $-0.006 \pm \; 0.0007$  & $-12.161 \pm \; 0.966$ & $-0.007 \pm \; 0.0006$ & $-2.558 \pm \; 0.184$ & $-0.006 \pm \; 0.0005$ \\
        Pf$_{15-5}$  & $-5.127 \pm \; 0.544$ & $-0.008 \pm \; 0.0008$  & $-12.426 \pm \; 0.95$ & $-0.007 \pm \; 0.0006$ & $-1.849 \pm \; 0.141$ & $-0.005 \pm \; 0.0004$ \\
        Pf$_{16-5}$  & $-4.288 \pm \; 0.412$ & $-0.006 \pm \; 0.0006$  & $-14.118 \pm \; 1.103$ & $-0.008 \pm \; 0.0007$ & $-1.796 \pm \; 0.132$ & $-0.004 \pm \; 0.0003$ \\
        Pf$_{17-5}$  & $-3.903 \pm \; 0.381$ & $-0.006 \pm \; 0.0006$  & $-10.712 \pm \; 0.832$ & $-0.006 \pm \; 0.0005$ & $-1.339 \pm \; 0.101$ & $-0.003 \pm \; 0.0003$ \\
        Pf$_{18-5}$  & $-3.326 \pm \; 0.327$ & $-0.005 \pm \; 0.0005$  & $-8.843 \pm \; 0.733$ & $-0.005 \pm \; 0.0005$ & $-1.117 \pm \; 0.095$ & $-0.003 \pm \; 0.0002$ \\
        Pf$_{19-5}$  & $-3.968 \pm \; 0.395$ & $-0.006 \pm \; 0.0006$  & $-9.012 \pm \; 0.716$ & $-0.005 \pm \; 0.0005$ & $-1.865 \pm \; 0.209$ & $-0.005 \pm \; 0.0005$ \\
        Pf$_{20-5}$  & $-2.631 \pm \; 0.251$ & $-0.004 \pm \; 0.0004$  & $-6.30 \pm \; 0.497$ & $-0.004 \pm \; 0.0003$ & $-0.894 \pm \; 0.273$ & $-0.002 \pm \; 0.0007$ \\
        Pf$_{21-5}$  & $-3.097 \pm \; 0.305$ & $-0.005 \pm \; 0.0005$  & $-6.73 \pm \; 0.513$ & $-0.004 \pm \; 0.0003$ & $-2.006 \pm \; 0.192$ & $-0.005 \pm \; 0.0005$ \\
        Pf$_{22-5}$  & $-2.479 \pm \; 0.242$ & $-0.004 \pm \; 0.0004$  & $-4.773 \pm \; 0.359$ & $-0.003 \pm \; 0.0002$ & $-0.940 \pm \; 0.135$ & $-0.002 \pm \; 0.0003$ \\

        \bottomrule
    \end{tabular}
\end{table*}

\begin{table*}[h]
    \centering
    \caption{Emission line properties for sources 469 and 823.}
    \label{tab:combined}
    \begin{tabular}{l c c c c c c}
        \toprule
        & \multicolumn{2}{c}{\textbf{Source 469}} & \multicolumn{2}{c}{\textbf{Source 823}} \\
        \cmidrule(lr){2-3} \cmidrule(lr){4-5}
        Parameter & EW (\AA) & Luminosity ($L_{\odot}$) & EW (\AA) & Luminosity ($L_{\odot}$) \\
        \midrule
        Pa$_\alpha$  & $-213.614 \pm \; 16.073$ & $-0.185 \pm \; 0.015$  & $-35.497 \pm \; 3.052$ & $-0.026 \pm \; 0.0024$ \\
        Br$_{6-4}$   & $-16.45 \pm \; 1.2587$   & $-0.029 \pm \; 0.0023$ & $-8.483 \pm \; 0.725$ & $-0.006 \pm \; 0.0006$ \\
        Br$_{7-4}$   & $-16.843 \pm \; 1.2705$  & $-0.022 \pm \; 0.0017$ & $-7.055 \pm \; 0.604$ & $-0.006 \pm \; 0.0005$ \\
        Br$_{8-4}$   & $--$                     & $--$                   & $-7.492 \pm \; 0.683$ & $-0.006 \pm \; 0.0006$ \\
        Br$_{9-4}$   & $-15.424 \pm \; 1.155$   & $-0.012 \pm \; 0.001$  & $-7.459 \pm \; 0.716$ & $-0.005 \pm \; 0.0005$ \\
        Br$_{10-4}$  & $-14.052 \pm \; 1.073$   & $-0.011 \pm \; 0.0009$ & $-6.377 \pm \; 0.573$ & $-0.005 \pm \; 0.0004$ \\
        Br$_{11-4}$  & $-10.918 \pm \; 0.872$   & $-0.007 \pm \; 0.0006$ & $-4.227 \pm \; 0.363$ & $-0.003 \pm \; 0.0003$ \\
        Pf$_{11-5}$  & $--$                     & $--$                   & $--$                  & $--$ \\
        Pf$_{12-5}$  & $--$                     & $--$                   & $-1.719 \pm \; 0.155$ & $-0.001 \pm \; 0.0001$ \\
        Pf$_{13-5}$  & $-1.288 \pm \; 0.102$    & $-0.002 \pm \; 0.0002$ & $-1.499 \pm \; 0.146$ & $-0.001 \pm \; 0.0001$ \\
        Pf$_{14-5}$  & $-1.449 \pm \; 0.120$    & $-0.003 \pm \; 0.0002$ & $-1.552 \pm \; 0.175$ & $-0.001 \pm \; 0.0001$ \\
        Pf$_{15-5}$  & $-2.091 \pm \; 0.171$    & $-0.004 \pm \; 0.0003$ & $-1.622 \pm \; 0.179$ & $-0.001 \pm \; 0.0001$ \\
        Pf$_{16-5}$  & $-1.213 \pm \; 0.103$    & $-0.002 \pm \; 0.0002$ & $-1.144 \pm \; 0.145$ & $-0.001 \pm \; 0.0001$ \\
        Pf$_{17-5}$  & $-1.338 \pm \; 0.125$    & $-0.002 \pm \; 0.0002$ & $-0.887 \pm \; 0.090$ & $-0.001 \pm \; 0.0001$ \\
        Pf$_{18-5}$  & $-0.639 \pm \; 0.066$    & $-0.001 \pm \; 0.0001$ & $-0.523 \pm \; 0.063$  & $-0.0004 \pm \; 0.0$ \\
        Pf$_{19-5}$  & $-1.291 \pm \; 0.125$    & $-0.002 \pm \; 0.0002$ & $-1.07 \pm \; 0.152$  & $-0.001 \pm \; 0.0001$ \\
        Pf$_{20-5}$  & $--$    & $--$ & $--$ & $--$ \\
        Pf$_{21-5}$  & $--$    & $--$ & $--$ & $--$ \\
        Pf$_{22-5}$  & $--$    & $--$ & $--$ & $--$ \\
        \bottomrule
    \end{tabular}
    \tablefoot{\small{Values are missing for some emission lines, as these lines were either not present or or could not be measured reliably for these sources.}}
\end{table*}

\newpage

\section{Line profiles} \label{line_profiles}
\begin{figure*}[h!]
    \centering
    \includegraphics[width=0.9\linewidth]{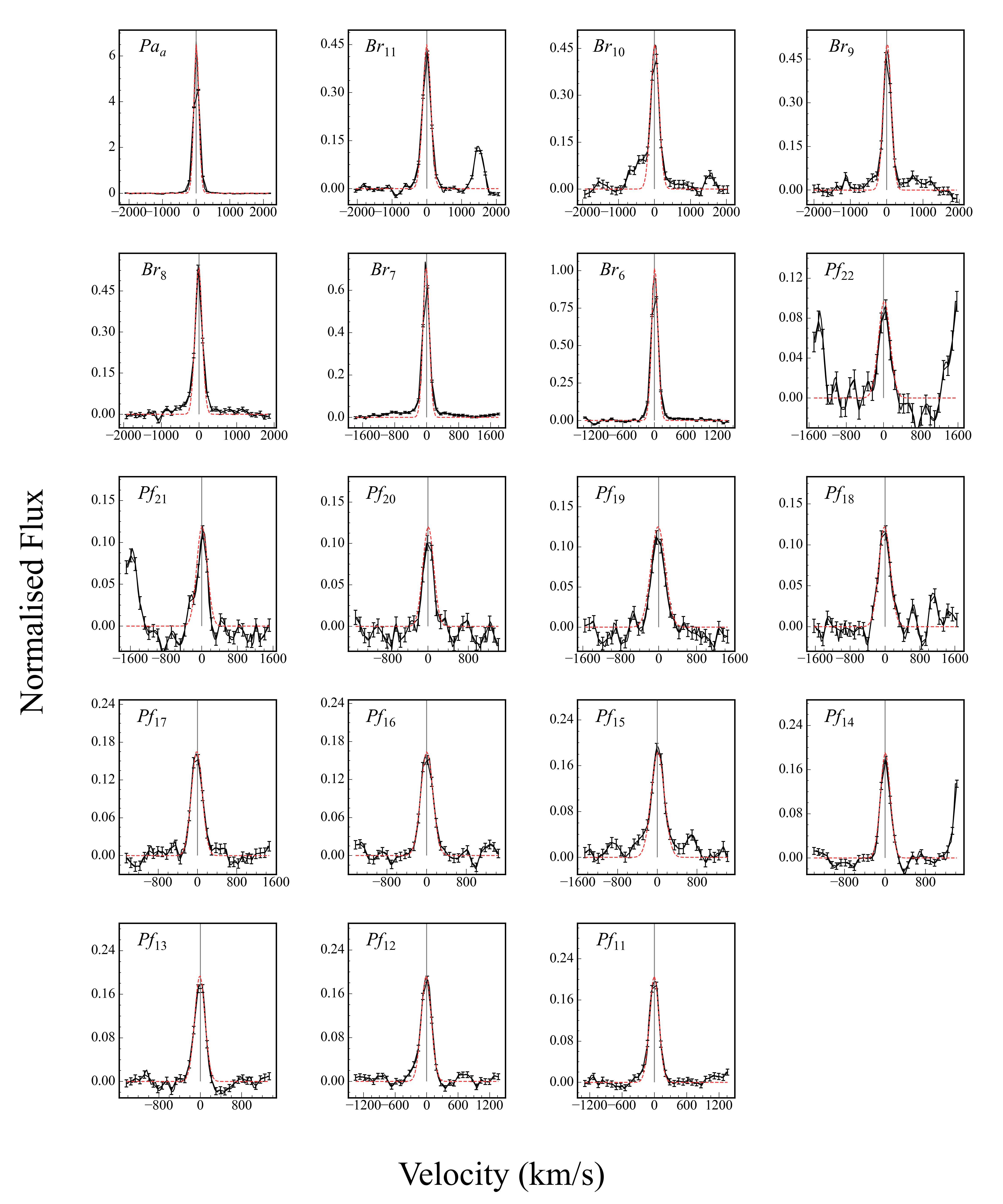}
    \label{fig:185_profiles}
    \caption{Line profiles for source 185.}
\end{figure*}

\begin{figure*}[h!]
    \centering
    \includegraphics[width=0.9\linewidth]{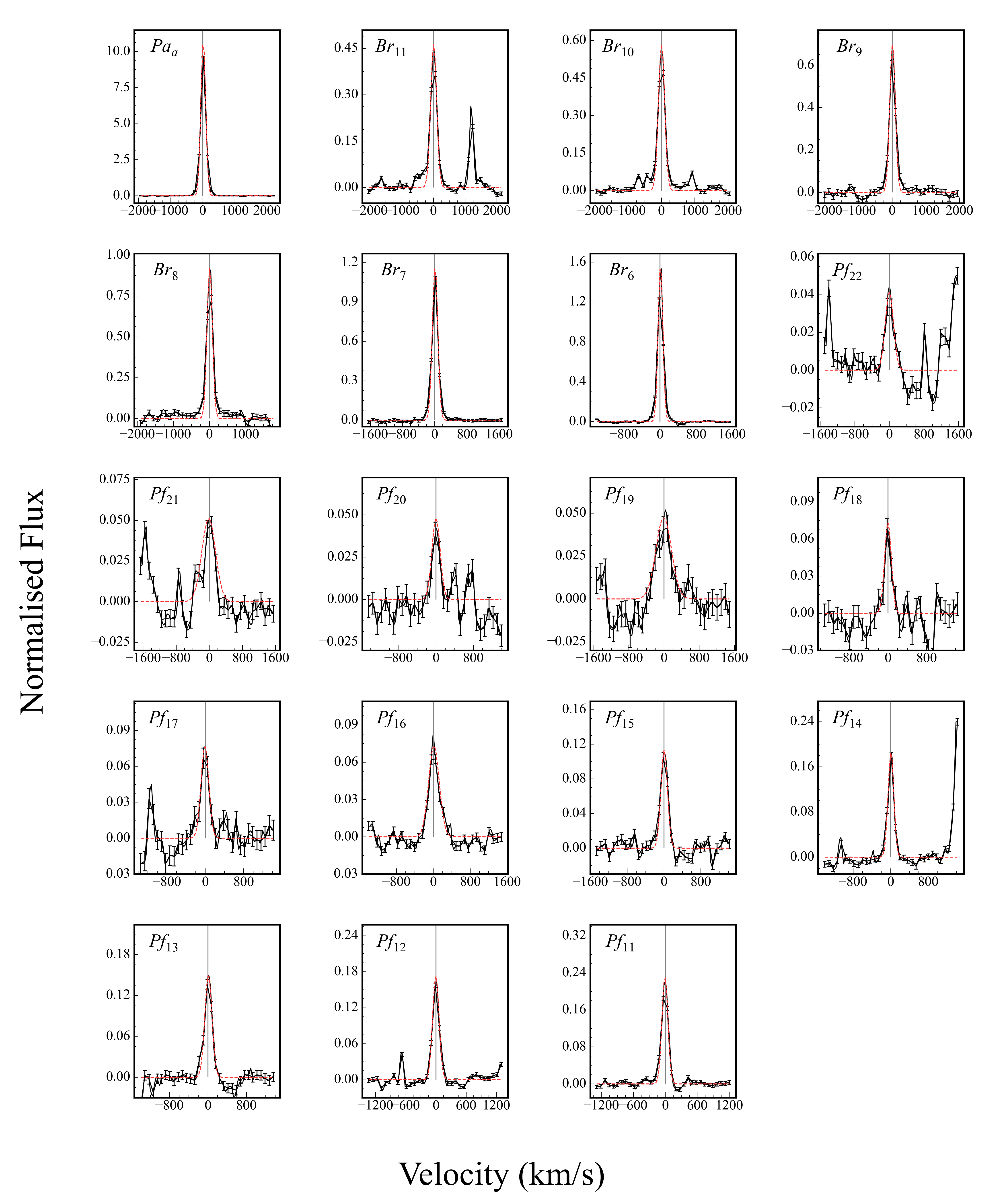}
    \label{fig:251_profiles}
    \caption{Line profiles for source 251. We note that $Pf_{22-5}$ and $Pf_{21-5}$ have been excluded due to low S/N.}
\end{figure*}

\begin{figure*}[h!]
    \centering
    \includegraphics[width=0.9\linewidth]{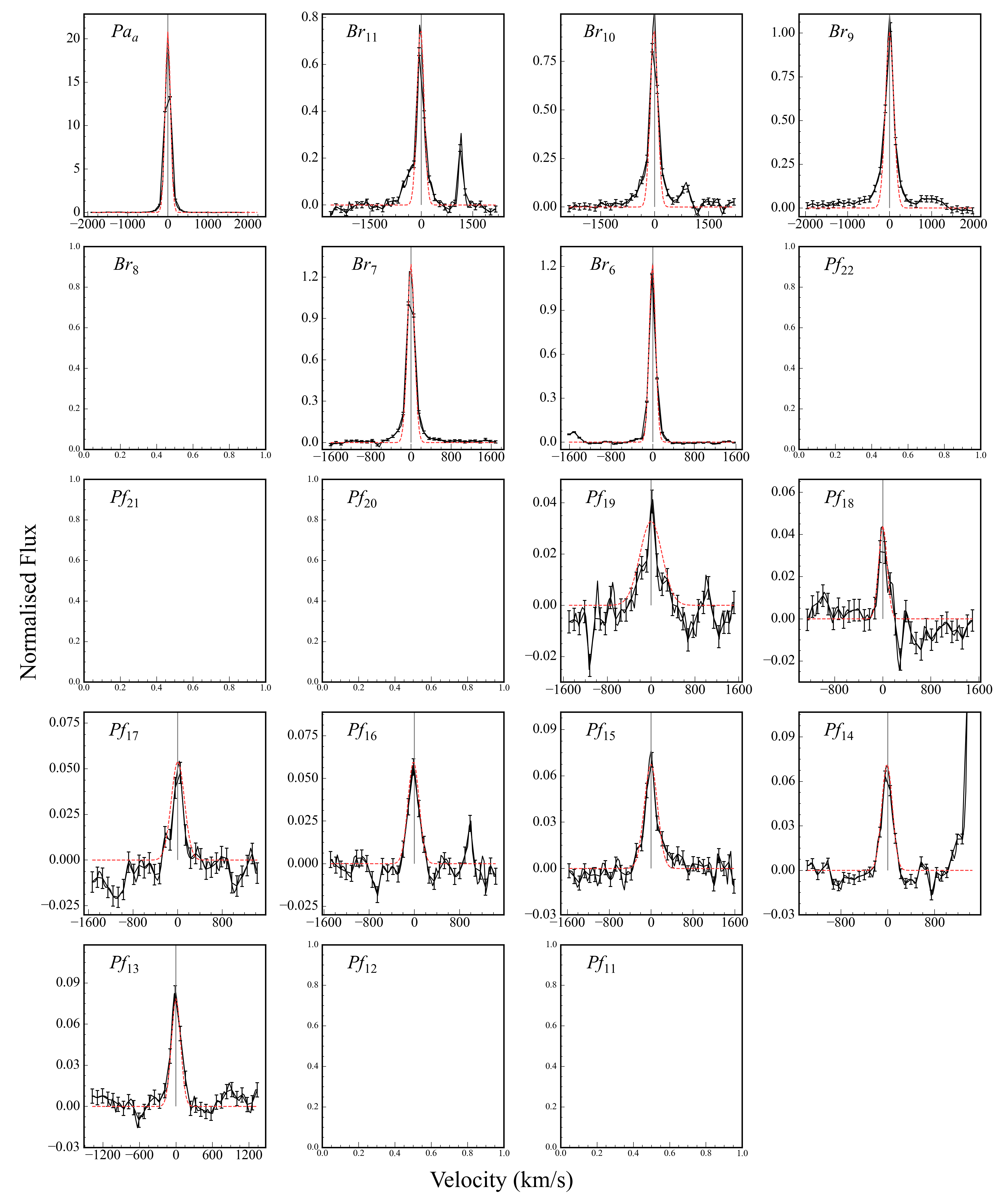}
    \label{fig:469_profiles}
    \caption{Line profiles for source 469. We note $Pf_{22}$, $Pf_{21}$, and $Pf_{20}$ have been excluded due to low S/N. Due to the location of the spectrum of source 469 on the NIRSpec detectors, $Br_{8}$, $Pf_{12}$, and $Pf_{11}$ were not measured on the detector.}
\end{figure*}

\begin{figure*}[h!]
    \centering
    \includegraphics[width=0.9\linewidth]{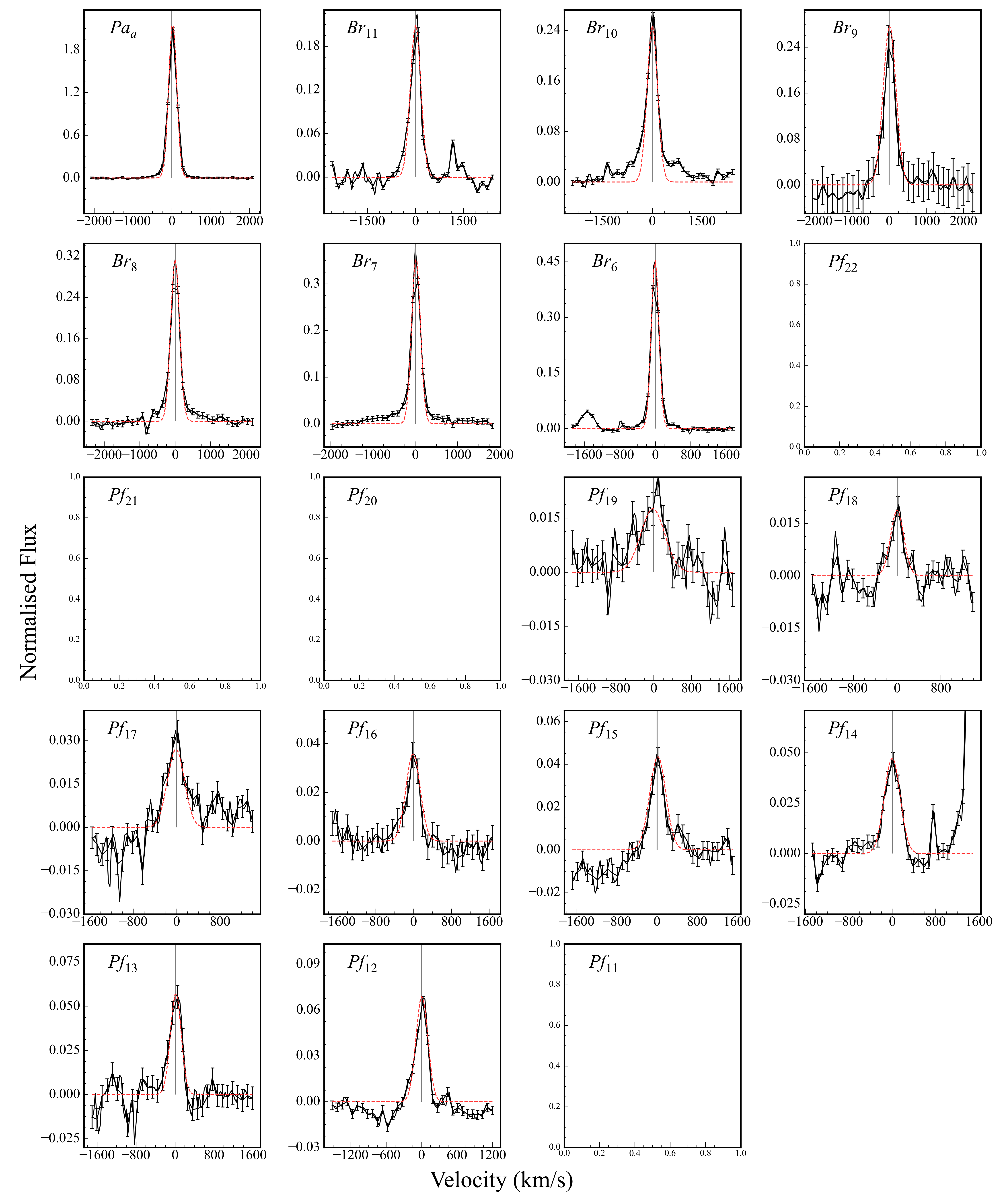}
    \label{fig:823_profiles}
    \caption{Line profiles for source 823. We note $Pf_{22}$, $Pf_{21}$, and $Pf_{20}$ have been excluded due to low S/N, and $Pf_{11}$ was not measured on the detector.}
\end{figure*}

\clearpage
 
\section{MCMC best-fit plots} \label{best_fitting_plots}
\begin{figure}[h!]
    \centering
    \includegraphics[width=0.65\linewidth]{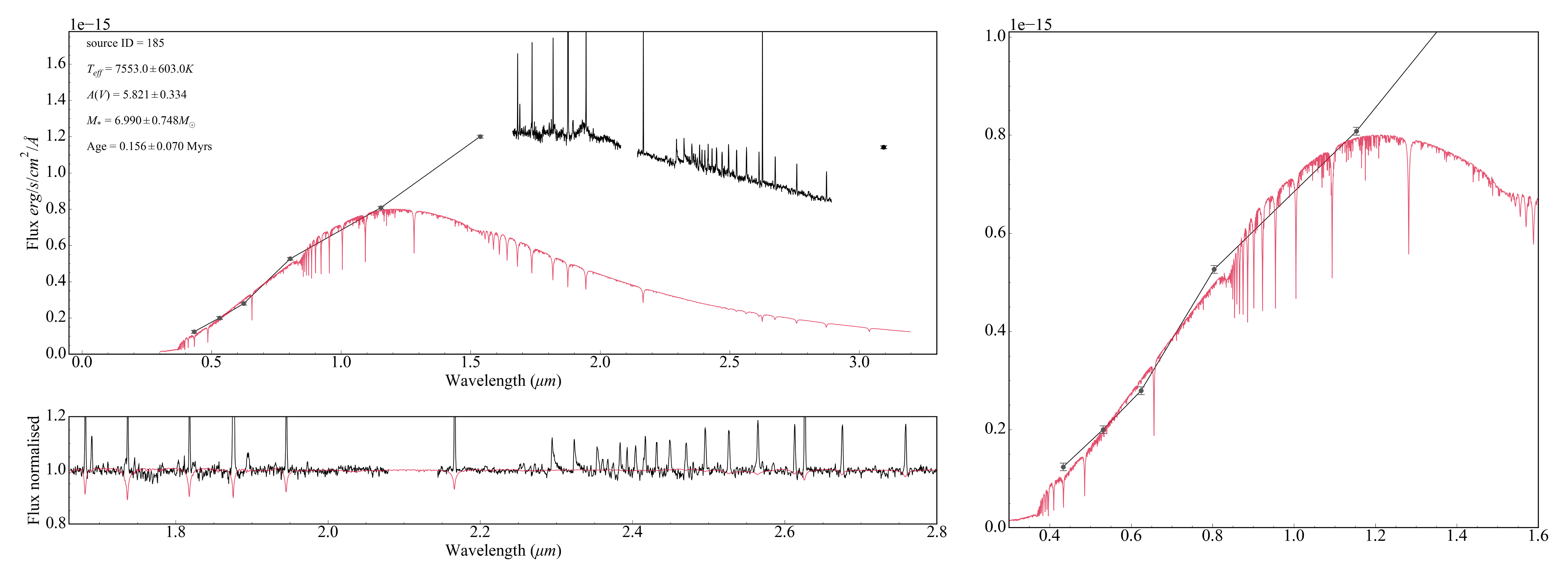}
    \label{fig:185_best_fit}
    \caption{Best-fitting spectrum for source 185.}
\end{figure}
\begin{figure}[h!]
    \centering
    \includegraphics[width=0.65\linewidth]{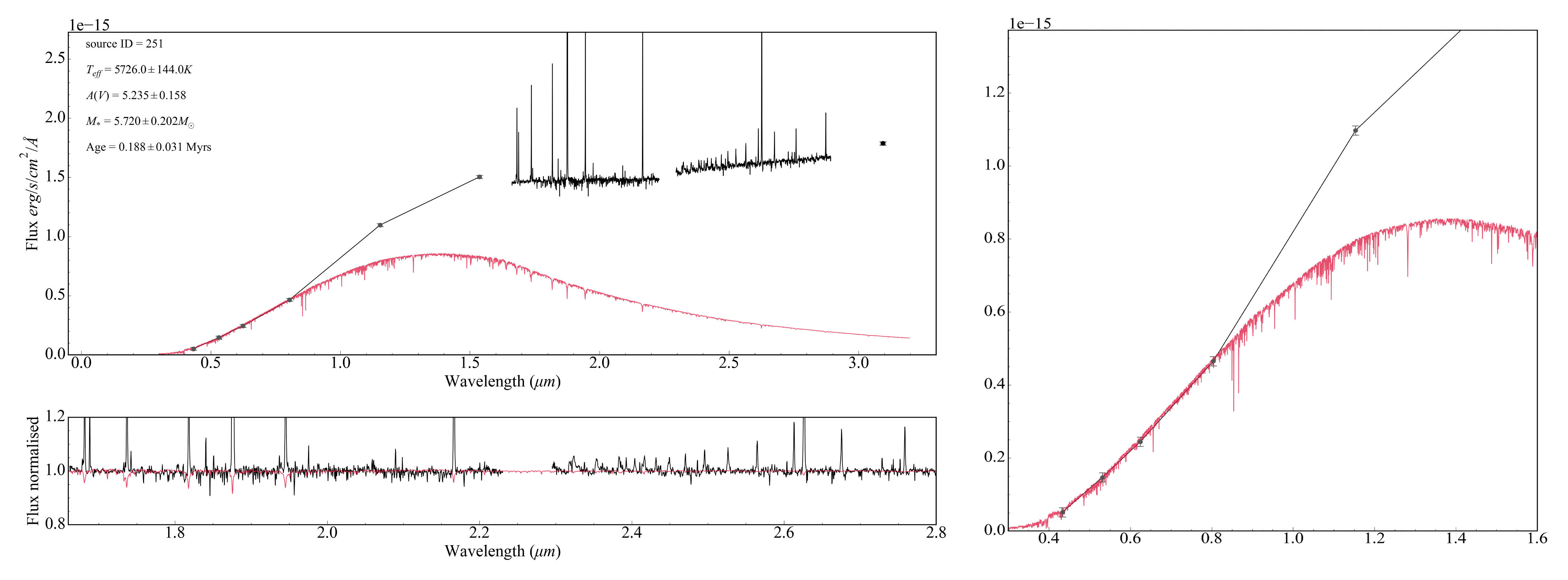}
    \label{fig:251_best_fit}
    \caption{Best-fitting spectrum for source 251.}
\end{figure}
\begin{figure}[h!]
    \centering
    \includegraphics[width=0.65\linewidth]{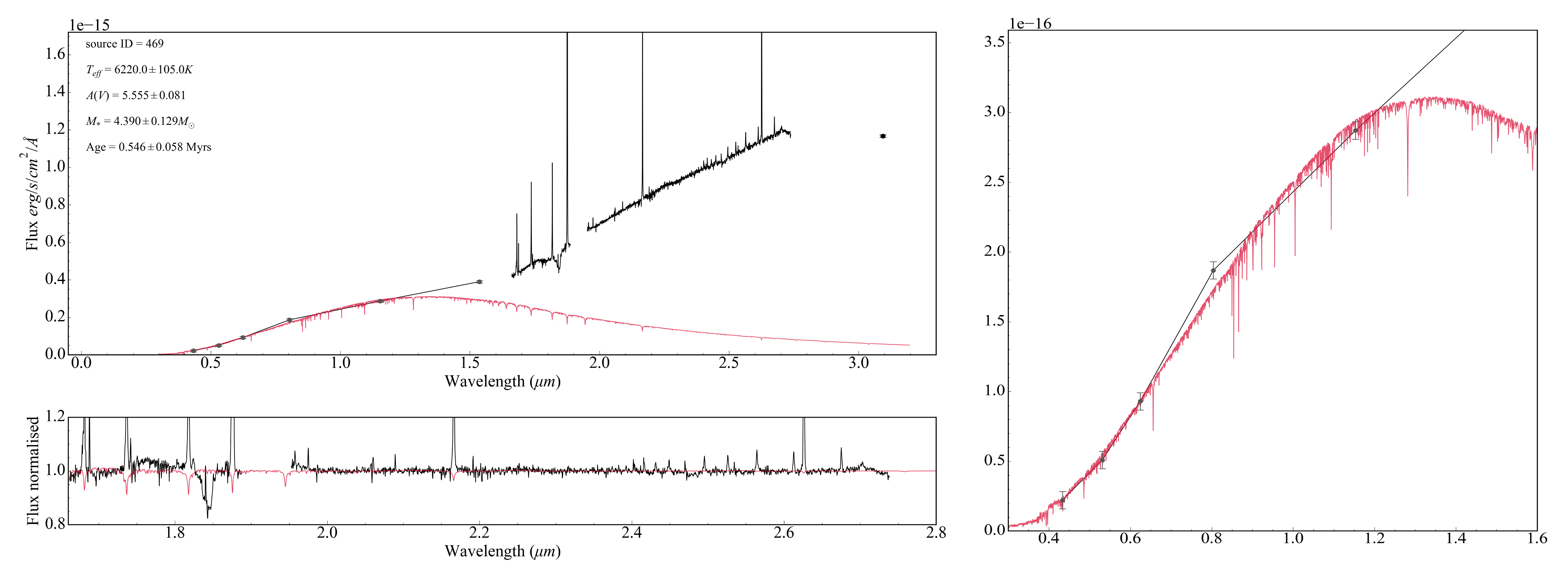}
    \label{fig:469_best_fit}
    \caption{Best-fitting spectrum for source 469.}
\end{figure}
\begin{figure}[h!]
    \centering
    \includegraphics[width=0.65\linewidth]{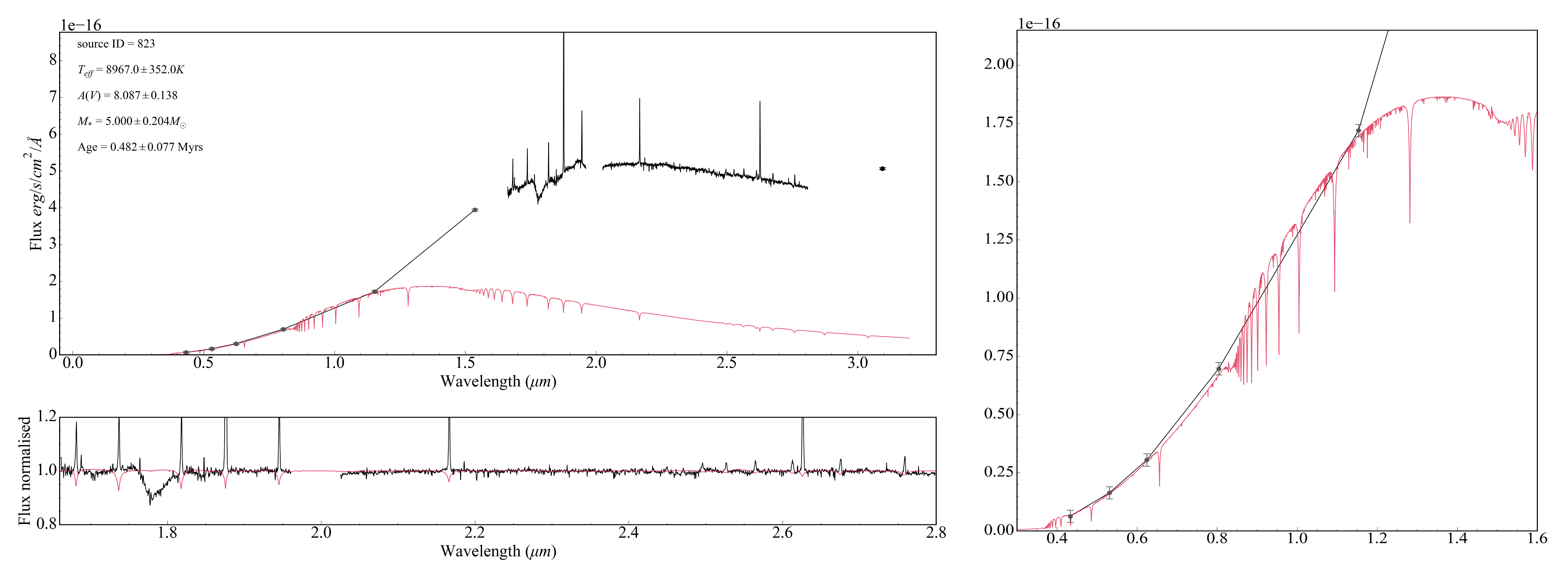}
    \label{fig:823_best_fit}
    \caption{Best-fitting spectrum for source 823.}
\end{figure}
\end{appendix}
\end{document}